\documentclass[aps,prb,reprint,superscriptaddress]{revtex4-2}
\usepackage{graphicx}
\usepackage{amsmath}
\usepackage{amssymb}
\usepackage{color}
\usepackage{hyperref}
\usepackage{siunitx}
\usepackage{chemformula}

\newcommand{\EMCP}{Eu$_2$CuMn$_2$P$_3$}
\newcommand{\ECZP}{Eu$_2$CuZn$_2$P$_3$}
\newcommand{\ECP}{EuCuP}
\newcommand{\EZP}{EuZn$_2$P$_2$}
\newcommand{\EMP}{EuMn$_2$P$_2$}
\newcommand{\C}{$^\circ$C}

\begin{document}

\title{Structural design and multiple magnetic orderings of the intergrowth compound \EMCP}
\author{Xiyu Chen}
\altaffiliation{The authors contributed equally to this work.}
\affiliation{Key Laboratory of Quantum Materials and Devices of Ministry of Education, School of Physics, Southeast University, Nanjing 211189, China}
\author{Ziwen Wang}
\altaffiliation{The authors contributed equally to this work.}
\affiliation{Key Laboratory of Quantum Materials and Devices of Ministry of Education, School of Physics, Southeast University, Nanjing 211189, China}
\author{Wuzhang Yang}
\affiliation{School of Science, Westlake University, Hangzhou 310024, China}
\affiliation{Institute of Natural Sciences, Westlake Institute for Advanced Study, Hangzhou 310024, China}
\author{Jia-Yi Lu}
\affiliation{School of Physics, Interdisciplinary Center for Quantum Information and State Key Laboratory of Silicon and Advanced Semiconductor Materials, Zhejiang University, Hangzhou 310058, China}
\author{Zhiyu Zhou}
\affiliation{Key Laboratory of Quantum Materials and Devices of Ministry of Education, School of Physics, Southeast University, Nanjing 211189, China}
\author{Zhi Ren}
\affiliation{School of Science, Westlake University, Hangzhou 310024, China}
\affiliation{Institute of Natural Sciences, Westlake Institute for Advanced Study, Hangzhou 310024, China}
\author{Guang-Han Cao}
\affiliation{School of Physics, Interdisciplinary Center for Quantum Information and State Key Laboratory of Silicon and Advanced Semiconductor Materials, Zhejiang University, Hangzhou 310058, China}
\affiliation{Collaborative Innovation Center of Advanced Microstructures, Nanjing University, Nanjing 210093, China}
\author{Shuai Dong}
\email{sdong@seu.edu.cn}
\affiliation{Key Laboratory of Quantum Materials and Devices of Ministry of Education, School of Physics, Southeast University, Nanjing 211189, China}
\author{Zhi-Cheng Wang}
\email{wzc@seu.edu.cn}
\affiliation{Key Laboratory of Quantum Materials and Devices of Ministry of Education, School of Physics, Southeast University, Nanjing 211189, China}
\date{\today}

\begin{abstract}

We report the design, synthesis, crystal structure, and physical properties of a layered intergrowth compound, \EMCP. The structure of \EMCP\ features an alternating arrangement of hexagonal \ECP\ block layers and trigonal \EMP\ block layers, interconnected through shared Eu planes. This structural hybridization leads to multiple magnetic orderings in \EMCP: weak antiferromagnetic (AFM) ordering of Mn at $T_\mathrm{N}^\mathrm{Mn}$ = 80 K, AFM ordering of Eu at $T_\mathrm{N}^\mathrm{Eu}$ = 29 K, a spin-reorientation transition at $T_\mathrm{SR}$ = 14.5 K, and weak ferromagnetism below $T_\mathrm{N}^\mathrm{Mn}$. The spin configurations at different temperature regions were discussed based on the calculations of magnetic energies for various collinear arrangements. Resistivity measurements reveal a pronounced transition peak at $T_\mathrm{N}^\mathrm{Eu}$, which is suppressed in the presence of a magnetic field, resulting in a significant negative magnetoresistance effect. The computed semimetallic band structure, characterized by a small density of states at the Fermi level, aligns well with experimental observations. The successful synthesis of \EMCP\ and its fascinating magnetic properties highlight the effectiveness of our block-layer design strategy. By assembling magnetic block layers of compounds with compatible crystal symmetries and closely matched lattice parameters, this approach opens exciting avenues for discovering layered materials with unique magnetic behaviors.

\end{abstract}

\maketitle

\section{Introduction}

Layered compounds represent a class of materials composed of two-dimensional (2D) slabs that are strongly bonded within each layer but are stacked with relatively weak interlayer forces~\cite{mckinney2018}. Their diverse structures and properties have spurred extensive research into harnessing and expanding their functionalities~\cite{yang2024a,willner1998,massidda1987,gascoin2005}. More importantly, the ability to replace or modify the building layers in these compounds offers significant opportunities to tailor new characteristics. This flexibility has opened up exciting avenues for material discovery, exemplified by the exploration of high-temperature superconductors~\cite{scalapino2012}. To date, hundreds of cuprates, nickelates, and iron-based superconductors have been reported, all retaining their superconductively-active motifs while incorporating different spacer layers~\cite{bednorz1986a,wu1987,tokura1990,park1995,kamihara2008,rotter2008,wang2008,hsu2008,li2019c,sun2023a,zhu2024}. However, identifying layered compounds with specific structures and features still involves considerable trial-and-error, primarily due to the complex structural and chemical challenges, as well as the lack of a comprehensive theoretical framework~\cite{chamorro2018,jain2013}.

To facilitate the discovery of multi-component, complex layered materials, we propose a block-layer (B-L) model for designing ionic intergrowth structures~\cite{wang2021e,wangzhi-cheng2018}. The intergrowth structure can be visualized as the alternating stacking of unit building block layers (BLs) from two simple layered compounds, with the BLs sharing a common layer. Two key design principles are theoretically formulated: the lattice mismatch between the constituent BLs should be minimized to reduce lattice deformation; charge transfer, driven by difference in chemical potential between the BLs, is essential for lowering the energy of the intergrowth structure~\cite{wang2021e}. The B-L model has proven its effectiveness through the discovery of dozens of iron-based superconductors~\cite{sun2012,liu2016,liu2016a,wang2016,wang2017,wang2017a,wu2017,shao2019a}. Our approach adopts a holistic mindset by using neutral BLs composed of multiple ionic layers as the fundamental building units, rather than relying on individual ionic layers. A key challenge in developing new intergrowth compounds lies in identifying layered materials that share similar crystal symmetries, have closely matched lattice parameters, and possess a common layer within both BLs. By addressing these criteria, we can systematically explore and create new layered materials.

In this work, we successfully designed and synthesized a novel quaternary phosphide, \EMCP, using a rational B-L design. The structure of \EMCP, illustrated in Fig. \ref{fig:STRUCTURE}, comprises alternating BLs from the hexagonal SrPtSb-type \ECP\ and the trigonal \ch{CaAl2Si2}-type \EMP, featuring closely matched in-plane lattice constants~\cite{tomuschat1981,ruhl1979}. To date, only two isostructural compounds have been reported: \ECZP\ and \ch{Ca2CuZn2P3}, first synthesized decades ago~\cite{frik1999}. These materials arise from the intergrowth of $A$CuP and $A$\ch{Zn2P2} ($A$ = Ca, Eu)~\cite{mewis1978a,tomuschat1981,klufers1977,klufers1980}. Our research is driven by three key motivations. First, to discover a new hybrid structure through top-down deterministic assembly rather than trial-and-error approaches. Second, to introduce novel magnetic interactions or modulate existing correlations via strategic B-L design. The Eu planes in \ECP\ exhibits ferromagnetic (FM) ordering at 32 K~\cite{iha2019,wang2023,May2023,yuan2024}. In contrast, the Eu sites in \EMP\ form antiferromagnetic (AFM) ordering at 17 K, while the Mn sites show only strong AFM correlations without long-range order in earlier studies~\cite{payne2002,berry2023}. However, a very recent study claims the presence of weak Mn-based ordering around 50 K~\cite{Krebber2025}. In the intergrowth structure \EMCP, the distance between neighboring Eu planes varies periodically, and the spacing between \ch{Mn2P2} layers increases significantly due to the intervening \ECP\ BLs. This altered magnetic environment imparts distinct properties to \EMCP, such as Mn ordering at 80 K, which is absent in \EMP. Third, to manipulate charge transport through B-L design. \ECP\ is metallic~\cite{iha2019,wang2023,May2023,yuan2024}, whereas \EMP\ is insulating~\cite{payne2002,berry2023}. \EMCP\ retains the conductivity of \ECP\ while preserving the structural integrity of the \EMP\ BLs. This B-L-design approach offers a novel method to tune transport properties without disrupting the individual layers within the BLs. The successful synthesis of \EMCP\ underscores the effectiveness of our B-L design strategy. It not only creates opportunities for discovering new layered materials, but also introduces unique magnetic behaviors and provides a means to manipulate transport properties.

\section{Methods}
\subsection{Crystal Growth}
Single crystals of \EMCP\ were grown using a Pb flux (99.999\%) from high-purity starting materials: Eu ingots (99.9\%), Mn powder (99.8\%), Cu powder (99.9\%), and P lumps (99.999\%). Prior to use, the surface of the Eu ingots was cleaned to remove any oxides and subsequently cut into small pieces. The elements were weighed with a molar ratio of Eu:Mn:Cu:P:Pb = 2:2:1:3:20, roughly mixed, and loaded into an alumina crucible that was placed within an evacuated quartz tube. The sealed tube was heated to 600\C\ over a period of 10 hours, held there for 5 hours, then further heated to 1100\C\ over another 10 hours and maintained at this temperature for 36 hours. Following this, the temperature was reduced to 800\C\ at a cooling rate of 2\C\ per hour, then to 700\C\ at a rate of 4\C\ per hour. Finally, the Pb flux was removed via centrifugation. The resulting \EMCP\ crystals appeared as lustrous hexagonal platelets with a typical dimension of $0.7\times0.7\times0.03$ mm$^3$ and demonstrated air stability.

The polycrystalline \EMCP\ samples were synthesized using conventional solid-state reactions. Stoichiometric amounts of high-purity Eu, Cu, Mn, and P (molar ratio 2:1:2:3) were roughly mixed and loaded into an alumina crucible, which was then sealed in an evacuated quartz tube. The sealed tube was heated to 900\C\ over 10 hours, maintained at this temperature for 30 hours, and subsequently cooled to room temperature naturally. The resulting product was reground and pressed into pellets. These pellets were placed in the alumina crucible and sealed again in an evacuated quartz tube. The sample was then heated to 1050\C\ and annealed for 60 hours, with an intermediate regrinding and pelleting step to ensure homogeneity. The characterization data for the polycrystalline sample are provided in the Supplemental Materials (SM)~\cite{suppmatt}, serving as complementary information to support the single-crystal \EMCP\ data presented in the main text.

\subsection{X-ray Diffraction}
Suitable crystals of \EMCP\ were selected for single-crystal x-ray diffraction (SCXRD) analysis. Data collection was carried out at 150 K using a Bruker D8 Venture diffractometer, which was equipped with an I$\mu$S 3.0 Dual Wavelength system (Mo $K\alpha$ radiation, $\lambda$ = 0.71073 \AA), and an APEXII CCD detector. The collected frames were reduced and corrected using the Bruker SAINT software suite. The initial structural model was developed with the intrinsic phasing feature of SHELXT and a least-square refinement was performed using SHELXL2014~\cite{Sheldrick}. Finally, the crystal structure was visualized using the VESTA software~\cite{VESTA}.

\subsection{Physical Measurements}
The direct-current (dc) magnetization was measured using a Quantum Design Magnetic Property Measurement System (MPMS-3) on a piece of \EMCP\ crystal with a total mass of about 0.1 mg. The alternating-current (ac) susceptibility was collected on a Physical Property Measurement System (PPMS-Dynacool, Quantum Design) with an ac measurement system (ACMS II). The electrical resistivity measurements were conducted on the PPMS-Dynacool utilizing a standard four-probe technique, and the electrodes were fabricated using gold wires attached with silver paste. The heat capacity was also measured using the PPMS-Dynacool with a relaxation time method on a collection of \EMCP\ crystals, totaling 2.09 mg in mass.

\subsection{Theoretical Calculations}

Density functional theory (DFT) calculations are performed using Vienna \textit{ab initio} Simulation Pack (VASP)~\cite{Kresse1996a}. The electron-ion interactions are described by projector augmented wave (PAW) pseudopotentials~\cite{Blochl1994a}. The plane-wave cutoff energy is fixed as 450 eV. The exchange-correlation functional is treated using Perdew-Burke-Ernzerhof (PBE) parametrization of the generalized gradient approximation (GGA)~\cite{Perdew1996}. A $\Gamma$-centered $11\times11\times2$ Monkhorst-Pack $k$-mesh is adopted for Brillouin zone sampling. To better describe the correlation effect, we apply an effective Hubbard $U$ value of 5.0 eV for the Eu 4$f$ orbitals and 2.0 eV for the Mn 3$d$ orbitals, using the Dudarev approach, as was done in earlier studies on Eu$^{2+}$-based systems~\cite{Dudarev1998,wangSinglePairWeyl2019,Chen2024a}. The lattice constants and atomic positions are optimized iteratively until the total energy and the Hellmann-Feynman force on each atom are converged to $10^{-6}$ eV and 0.02 eV/\AA, respectively.

In our calculations, we tested various $U_\mathrm{Mn}$ values for the Mn 3$d$ orbitals. While a larger $U_\mathrm{Mn}$ of 5 eV was used in previous studies on \EMP~\cite{berry2023}, our tests show that moderate values (2.0 eV and 3.0 eV) provide a more accurate description of both the lattice constants and the electronic structure when compared to our experimental data. Therefore, we present the results obtained with $U_\mathrm{Mn}$ = 2.0 eV in the main text. The band structures for $U_\mathrm{Mn}$ = 3.0 eV and 5.0 eV are detailed in Fig. S8 of the SM~\cite{suppmatt}.

\section{Results and discussion}

\subsection{Structural Analysis}

\begin{figure}
	\includegraphics[width=0.5\textwidth]{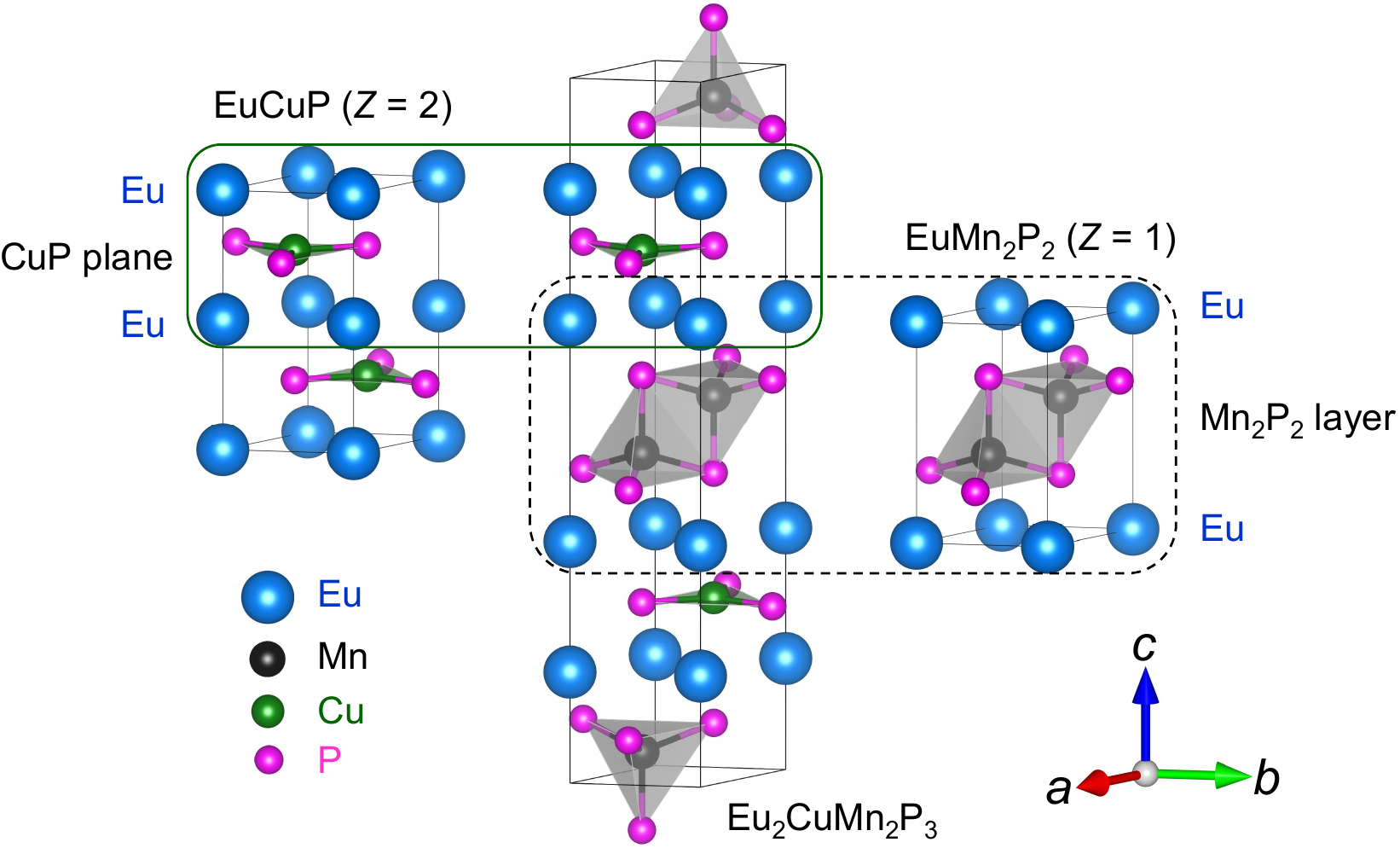}
	\caption{\label{fig:STRUCTURE}
		Crystal structures of \ECP\ (left), \EMCP\ (middle), and \EMP\ (right). The \ECP\ and \EMP\ BLs within \EMCP\ are highlighted with solid and dashed boxes, respectively.}
\end{figure}

\begin{table}
  \caption{Crystallographic parameters at 150 K and structure refinement statistics summarized for a single crystal of \EMCP~\cite{ccdcEuMnCuP}.}
  \begin{ruledtabular}\label{tbl:CRYSTALDATA}
  \begin{tabular}{lc}
	    compound  & \EMCP \\
	    \hline
	    crystal system  & hexagonal \\	
	    space group  & $P6_3/mmc$ (No. 194)  \\   
	    $a$ (\AA)  & 4.1207(2) \\  
	    $c$ (\AA)  &  22.2676(12) \\
	    $V$ (\AA$^3$)  & 327.45(4) \\                           
	    $Z$  & 2 \\
	    $\rho_\mathrm{calc}$ (g/cm$^3$)  &  5.784 \\
	    temperature (K)  & 150 \\
	    radiation  & Mo $K\alpha$ ($\lambda$ = 0.71073)\\
	    2$\theta$ range ($^\circ$) & 7.32 $\rightarrow$ 53.258 \\
	    $h$ & $-5 \rightarrow  5$\\
	    $k$ & $-5   \rightarrow   5$ \\
	    $l$ & $-28   \rightarrow  28$ \\
	    reflections collected  & 4369 \\ 
	    independent reflections  & 171 \\
	    $R\rm_{int}$ & 0.0713 \\
	    Goodness-of-fit  & 1.230 \\
	    $R_1$\footnote{$R_1=\Sigma||F_o|-|F_c||/\Sigma|F_o|$.} [$F^2>2\sigma(F^2)$]  & 0.0333 \\
	    $wR_2$\footnote{$wR_2=[\Sigma w(F_o^2-F_c^2)^2/\Sigma w(F_o^2)^2]^{1/2}$.} [$F^2$]    & 0.1076 \\
	  \end{tabular}
	\end{ruledtabular}
\end{table}

\begin{table*}
	\caption{Fractional atomic coordinates, site occupancies, anisotropic displacement parameters ($U_{ij}$), and equivalent isotropic displacement parameters ($U_\mathrm{eq}$) for each Wyckoff site in the structure of \EMCP.}
	\begin{ruledtabular}\label{tbl:COORDINATES}
	\begin{tabular}{lllllllll}
		atom  & site & $x$ & $y$ & $z$ & occ. & $U_{11}$ (\AA$^2$)\footnote{$U_{22}=U_{11}$, $U_{12}=0.5U_{11}$, and $U_{13}=U_{23}=0$.} & $U_{33}$ (\AA$^2$) & $U_\mathrm{eq}$ (\AA$^2$)\footnote{$U_\mathrm{eq}$ is defined as 1/3 of of the trace of the orthogonalized $U_{ij}$ tensor.} \\
		\hline
		Eu & $4e$ & 0 & 0  & 0.15749(3) & 1.00 & 0.0187(6) & 0.0195(9) & 0.0190(6)\\
		Mn & $4f$ & 1/3 & 2/3 & 0.53824(11) & 1.00 & 0.0191(8) & 0.0219(16) & 0.0201(7)\\
		Cu & $2c$ & 1/3 & 2/3 & 1/4 & 1.00 & 0.0193(9) & 0.0228(18) & 0.0205(8)\\
		P1 & $2d$ & 1/3 & 2/3 & 3/4 & 1.00 & 0.0197(17) & 0.024(3) & 0.0210(14)\\
		P2 & $4f$ & 1/3 & 2/3 & 0.0727(2) & 1.00 & 0.0196(11) & 0.025(3) & 0.0214(10)\\
	\end{tabular}
\end{ruledtabular}
	
\end{table*}

The crystal structure of \EMCP\ is built up by alternately stacking the half-unit cell of \ECP\ ($Z=2$) and the complete unit cell of \EMP\ ($Z=1$), as illustrated in Fig. \ref{fig:STRUCTURE}. \ECP\ crystallizes in the hexagonal SrPtSb-type structure ($P6_3/mmc$ space group, No. 194), composed of honeycomb CuP planes separated by a triangular Eu$^{2+}$ lattice~\cite{tomuschat1981}. \EMP\ forms in the trigonal \ch{CaAl2Si2}-type structure ($P\bar{3}m1$ space group, No. 164), comprising layers of Eu$^{2+}$ ions sandwiched by layers of edge-sharing \ch{MnP4} tetrahedra~\cite{ruhl1979}. Both compounds conform to the Zintl concept, and the cations can be considered to have the valences Eu$^{2+}$, Cu$^{1+}$, and Mn$^{2+}$. The $a$ lattice constant at ambient temperature for \ECP\ (4.1205 \AA) is only 0.2\% smaller than that for \EMP\ (4.1294 \AA), indicating that the lattice deformation of the two BLs will be minimal when forming the intergrowth phase. Given the similar crystal symmetry and close in-plane lattice constants, the intergrowth phase \EMCP\ can be designed by inserting the \EMP\ BL into \ECP\ through the sharing of the Eu$^{2+}$ plane.

Single crystals of \EMCP\ are successfully prepared using a Pb-flux method. The crystallographic data at 150 K and refinement result of \EMCP\ from single-crystal x-ray diffraction (XRD) are listed in Tables \ref{tbl:CRYSTALDATA} and \ref{tbl:COORDINATES}, and the powder XRD pattern at room temperature is provided in the SM (Fig. S3)~\cite{suppmatt}. The resultant $a$ lattice constant of \EMCP\ (4.1207 \AA\ at 150 K, 4.127 \AA\ at room temperature) lies between those of \ECP\ and \EMP, while the $c$ lattice constant (22.2676 \AA\ at 150 K, 22.29 \AA\ at room temperature) is noticeably larger than, yet very close to, the sum of that of \ECP\ (8.1939 \AA) and twice of that of \EMP\ ($2\times6.9936$ \AA). The expansion of $c$ is attributed to the elastic deformations of the BLs and charge redistribution between them. Moreover, the sum of the unit cell volumes of \ECP\ and \EMP\ ($V_{111}+2V_{122}$) is nearly identical to the volume of \EMCP\ (327.45 \AA$^3$). The reasonable lattice constants and unit cell volume validate that the designed structure of \EMCP\ shown in Fig. \ref{fig:STRUCTURE} has been successfully synthesized. It is worth noting that the site mixing of Mn$^{2+}$ and Cu$^{1+}$ is not observed in our refinement, despite both being 3$d$ transition metal elements. This is plausible due to the distinct differences in their bonding environments, ionic radii, and chemical valences.

\subsection{Magnetic Properties}

\begin{figure*}
	\includegraphics[width=1\textwidth]{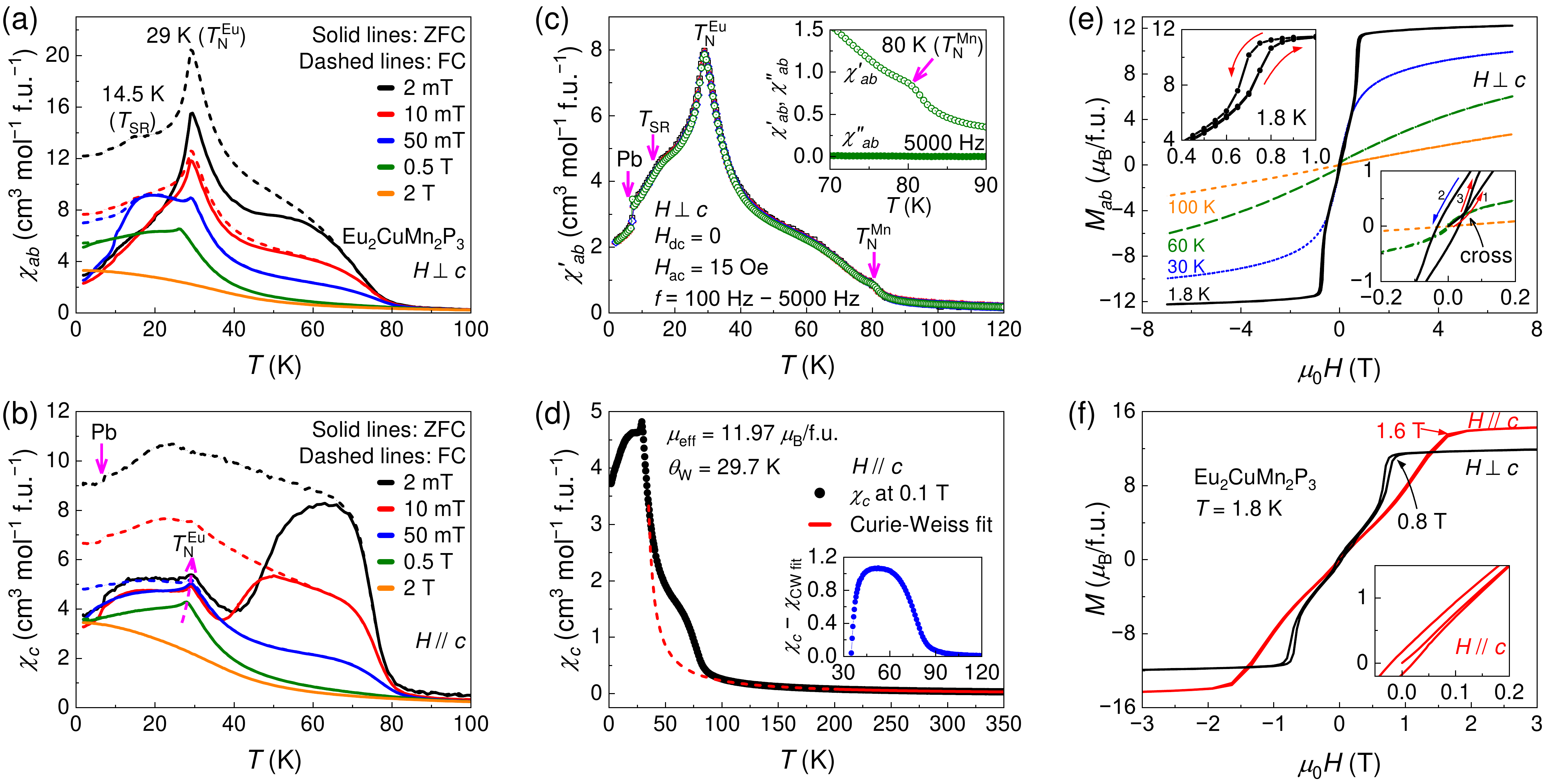}
	\caption{\label{fig:SUSCEPTIBILITY}
		Magnetic behaviors of \EMCP. (a) Temperature dependence of magnetic susceptibility, $\chi_{ab}(T)$, under various in-plane fields. The ZFC and FC data are displayed as solid and dashed lines, respectively. (b) Temperature dependence of susceptibility along the $c$ axis, $\chi_c(T)$. The magenta arrow around 7 K highlights the superconducting signature caused by residual Pb flux on the sample surface. (c) Temperature dependence of the in-phase component of ac magnetic susceptibility, $\chi^\prime_{ab}(T)$, at frequencies of 100 Hz (black), 368 Hz (red), 1366 Hz (blue), and 5000 Hz (olive). The inset shows a zoomed-in view of both the in-phase and out-of-phase components around 80 K, highlighting the transition. (d) Curie--Weiss analysis from 200 K to 350 K using $\chi_c(T)$ data at 0.1 T. The inset shows the discrepancy between $\chi_c(T)$ and the Curie--Weiss fit below 120 K.  (e) In-plane magnetization, $M_{ab}$, as a function of field at several temperatures. Detailed behaviors are shown in the two insets. (f) Comparison of the in-plane and out-of-plane magnetization at 1.8 K. The lower-right inset shows the magnetic hysteresis at the low fields.}
\end{figure*}

\EMCP\ exhibits a hybrid structure comprising BLs from \ECP\ and \EMP, and therefore displays magnetic characteristics that are both related to and distinct from those of its constituent phases. Previous studies have shown that \ECP\ is a ferromagnet with a Curie temperature ($T_\mathrm{C}$) of 32 K~\cite{iha2019,wang2023,May2023}, while \EMP\ displays A-type AFM order below the N\'{e}el temperature ($T_\mathrm{N}$) of 17 K for the Eu sites, with no detectable Mn ordered moment~\cite{payne2002,berry2023,Krebber2025}. The magnetic properties of \EMCP\ are summarized in Fig. \ref{fig:SUSCEPTIBILITY}. As shown in panel (a), the in-plane susceptibility ($\chi_{ab}$) increases markedly below 80 K and exhibits a bifurcation between zero-field-cooling (ZFC) and field-cooling (FC) measurements under a small field of 2 mT. The rapid increase in susceptibility and the divergence between ZFC and FC data typically indicate the presence of a FM contribution. The FM-like behavior is reproducible across multiple single-crystal \EMCP\ specimens; therefore, it cannot be simply attributed to impurity phases or experimental artifacts. However, in Fig. \ref{fig:SUSCEPTIBILITY}(c), the ac susceptibility reveals only a subtle peak around 80 K for the in-phase component $\chi^\prime_{ab}$, with no notable out-of-phase response $\chi^{\prime\prime}_{ab}$ (see the inset). Furthermore, the $\chi^\prime_{ab}$ peak exhibits no measurable frequency dependence as the frequency increases from 100 Hz to 5000 Hz, indicating that the dominant magnetic order observed at 80 K is AFM, whether long-range or short-range in nature. We designate this temperature as $T_\mathrm{N}^\mathrm{Mn}$, marking the characteristic transition point.

Below 40 K, $\chi_{ab}$ (Fig. \ref{fig:SUSCEPTIBILITY}(a)) and $\chi^\prime_{ab}$ (Fig. \ref{fig:SUSCEPTIBILITY}(c)) increase conspicuously as the temperature decreases, followed by a rapid decline below 29 K. As shown in Fig. \ref{fig:SUSCEPTIBILITY}(a), the transition peak at 29 K shifts to lower temperatures with stronger fields, accompanied by a gradually diminishing split between ZFC and FC curves. Additionally, a small kink is observed at 14.5 K in the FC $\chi_{ab}(T)$ curve at 2 mT, as well as in the $\chi^\prime_{ab}(T)$ curves. Compared to $\chi_{ab}(T)$ data, the out-of-plane susceptibility ($\chi_c$) in Fig. \ref{fig:SUSCEPTIBILITY}(b) also shows a significant rise below $T_\mathrm{N}^\mathrm{Mn}$. However, the upturn below 40 K and the transition peak at 29 K are negligible for the FC curves at 2 and 10 mT. For the curves with higher fields, the magnetic anisotropy between $\chi_{ab}(T)$ and $\chi_c(T)$ is relatively minor.

Given that the Eu lattices in \ECP\ and \EMP\ order at 32 K and 17 K, respectively, and considering that the exchange interactions within the BLs of \EMCP\ should not differ significantly from those in the respective compounds, it is reasonable to attribute the peak behavior at 29 K and the kink at 14.5 K to the ordering of the Eu lattice. The peak at 29 K results from the AFM ordering, with the easy magnetization direction lying in the $ab$ plane. This conclusion is supported by the following observations: (i) both the FC and ZFC $\chi_{ab}$ curves exhibit a sharp decline below the transition temperature; (ii) the transition peak shifts to lower temperatures under higher magnetic fields; (iii) the fact that $\chi_c$ decreases mildly below the peak, while $\chi_{ab}$ decreases sharply, indicates that the $ab$ plane is the magnetic easy plane. The easy-plane AFM configuration of \EMCP\ is further corroborated by the metamagnetic transition observed when the field is applied in the $ab$ plane, as illustrated in Fig. \ref{fig:SUSCEPTIBILITY}(f). The Eu spin configuration of \EMCP\ resembles the magnetic structures of \EMP\ and other Eu-based layered compounds, characterized by the intralayer FM and interlayer AFM couplings~\cite{chen2024}. However, the varying magnetic exchanges across the \ECP\ and \EMP\ BLs introduce additional complexity to \EMCP: the overall interlayer AFM coupling can be achieved through a variety of exchange interaction combinations. Moreover, the kink observed at 14.5 K is likely indicative of a spin-reorientation transition within the Eu plane, possibly driven by the competition between exchange interactions and magnetocrystalline anisotropy at low temperatures. The temperatures of the peak and the kink are designated as $T_\mathrm{N}^\mathrm{Eu}$ and $T_\mathrm{SR}$, respectively. A detailed discussion of the spin configuration of \EMCP\ will follow, incorporating magnetic energy calculations.

The AFM-like transition at $T_\mathrm{N}^\mathrm{Mn}$ and the emergence of the FM contribution below $T_\mathrm{N}^\mathrm{Mn}$ cannot be owed to the ordering of Eu spins, as this would conflict with the AFM transition at $T_\mathrm{N}^\mathrm{Eu}$. Consequently, Mn$^{2+}$ emerges as the sole candidate responsible for the transition at $T_\mathrm{N}^\mathrm{Mn}$, given that Cu$^{1+}$ is nonmagnetic. Although long-range Mn order is absent in \EMP\ in some previous studies, Mn magnetic correlations have been observed to develop~\cite{payne2002,berry2023}. This lack of long-range order is attributed to the chemical pressure in \EMP~\cite{berry2023}, which may be significantly altered in the intergrowth structure of \EMCP. Moreover, the sibling compounds \ch{SrMn2P2} and \ch{EuMn2As2} exhibit long-range AFM Mn order at 53 K and 142 K, respectively~\cite{sangeetha2021,anand2016,dahal2019}. Therefore, it is reasonable to infer that the magnetic ordering at $T_\mathrm{N}^\mathrm{Mn}$ and the weak ferromagnetism below $T_\mathrm{N}^\mathrm{Mn}$ are linked to the Mn sites. The FM contribution to the susceptibility is essentially saturated below 60 K, as indicated by the plateaus in the $\chi_{ab}(T)$ and $\chi_c(T)$ curves around 60 K. Figure \ref{fig:SUSCEPTIBILITY}(d) shows the results of a Curie--Weiss analysis, which we use to extract the contribution from the weak FM state. To avoid potential influences from magnetic fluctuations, we select and fit the $\chi_c(T)$ data at 0.1 T from 200 to 350 K to the Curie--Weiss law $\chi(T) = C/(T-\theta)+\chi_0$. This yields a Weiss temperature of $\theta=29.7$ K, an effective moment of $\mu_\mathrm{eff}=11.97$ $\mu_B$/f.u. ($C=17.9$ K cm$^3$ mol$^{-1}$ f.u.$^{-1}$), and $\chi_0=-0.02571$ cm$^3$ mol$^{-1}$ f.u.$^{-1}$. The positive $\theta$ suggests that the intralayer Eu-Eu FM interaction is dominant, while the small negative $\chi_0$ stems from the signal of low-temperature varnish. Notably, $\mu_\mathrm{eff}$ closely matches the theoretical value of 11.2 $\mu_B$/f.u. ($7.94\sqrt{2}$) for two Eu$^{2+}$ ions, which is considerably lower than the sum of the free ion values of 14.0 $\mu_B$/f.u. ($\sqrt{2\times7.94^2+2\times5.92^2}$) for two Eu$^{2+}$ and two Mn$^{2+}$ ions. The absence of Mn contribution to the paramagnetic (PM) susceptibility is a common feature in \ch{CaAl2Si2}-type Mn-based compounds, attributed to strong AFM correlations well above the ordering temperature~\cite{berry2023,anand2016,sangeetha2021,sangeetha2016}.

The calculated curve using the fitted parameters of the Curie--Weiss analysis is compared to $\chi_c(T)$ in Fig. \ref{fig:SUSCEPTIBILITY}(d), and the discrepancy is shown in the inset, with a maximum around 55 K. Consequently, we conclude that the susceptibility contributed by the FM state ($\chi_c-\chi_\mathrm{CW\,fit}$) is approximately 1.05 cm$^3$ mol$^{-1}$ f.u.$^{-1}$ under 0.1 T and the saturated moment is calculated to be 0.19 $\mu_B$/f.u. ($1.05\times10^3~\mathrm{emu~mol}^{-1}=0.19~\mu_B/\mathrm{f.u.}\sim0.09 \mu_B/\mathrm{Mn}$), which is much smaller than the local moment of each Mn$^{2+}$ ($S=5/2$, and $\mu=4.1\ \mu_B$ from the subsequent theoretical calculations). The saturated moment of the FM component, derived from susceptibility analysis, agrees well with the value obtained by linearly extrapolating the $c$-axis $M(H)$ curve at 60 K, as shown in Fig. S2(d) of the SM~\cite{suppmatt}. The weak ferromagnetism of \EMCP\ can be explained by either the local or itinerant scenarios. In the local scenario, a slight spin canting of the Mn local moment accounts for the FM component. In the itinerant scenario, itinerant carriers in the lattice are spin-polarized by the AFM Mn local moment, leading to the weak FM state. The hole concentration of \EMCP, estimated from Hall resistivity measurements, is $4.3 \times 10^{20}$~cm$^{-3}$ (Fig.~\ref{fig:RESISTIVITY}(c)). Assuming all hole carriers order ferromagnetically, this yields a saturation magnetization of $\mu_{\mathrm{FM}} = n g S \mu_{\mathrm{B}} = 0.07~\mu_{\mathrm{B}}/\mathrm{f.u.}$, where $g \approx 2$, $S = 1/2$, and $n$ is the hole density per formula unit. This theoretical estimate is reasonably consistent with the experimental saturation moment, considering the uncertainties in both the experimental value estimation and the hole concentration determination. Both scenarios are compatible with the AFM ordering of Mn moments at $T_\mathrm{N}^\mathrm{Mn}$. Furthermore, the observation of a weak transition peak associated with AFM Mn ordering is not unexpected, given the quasi-2D nature of Mn correlations in \EMCP~\cite{T_N_2}, which arises from the substantially larger inter-B-L Mn-Mn distance compared to that in \ch{CaAl2Si2}-type Mn-based materials~\cite{ruhl1979,anand2016,mewis1978}. A Curie--Weiss analysis of $\chi_{ab}(T)$ is also presented in Fig. S2(a), yielding results similar to those in Fig. \ref{fig:SUSCEPTIBILITY}(d).

The magnetization isothermal curves with the in-plane fields $M_{ab}(H)$ are presented in Fig. \ref{fig:SUSCEPTIBILITY}(e). At 100 K, the orange curve is a perfect linear line, while the olive curve at 60 K shows a rapid increase in the low-field region ($\mu_0H<50$ mT), attributed to the weak FM state below $T_\mathrm{N}^\mathrm{Mn}$. Magnetic hysteresis has been observed at this temperature, as specifically shown in Fig. S2(d). The blue curve at 30 K, near the AFM transition temperature $T_\mathrm{N}^\mathrm{Eu}$, follows a Brillouin function due to the PM contribution of Eu$^{2+}$. The black curve at 2 K reveals unique magnetic characteristics. First, a distinct hysteresis loop with a coercivity of 30 mT is observed, as shown in the magnified view in the bottom-right inset. Notably, the initial branch (marked as 1) intersects with the final field-increasing branch (marked as 3), a phenomenon rarely seen in typical hysteresis loops. This behavior is reproducible for in-plane magnetization but absent in the out-of-plane case, as illustrated in the inset in Fig. \ref{fig:SUSCEPTIBILITY}(f). The observed feature may be attributed to domain movement in easy-plane A-type AFM systems, as discussed in ref.~\cite{pakhira2022a}. Second, a step-like increase in magnetization due to spin rotation is observed at 0.6 T, as shown in the upper-left inset. With further increases in the magnetic field, the hysteresis loop reopens. Enhanced hysteresis after the metamagnetic transition has also been reported in other materials with coexisting metamagnetic transitions and FM components, possibly due to significant changes in the internal field caused by flipped spins~\cite{wang2021a}. The comparison between $M_{ab}(H)$ and $M_c(H)$ curves at 1.8 K is shown in Fig. \ref{fig:SUSCEPTIBILITY}(f). The saturation field ($H_\mathrm{sat}$) for $M_c(H)$ (1.6 T) is twice as that for $M_{ab}(H)$ (0.8 T), and the metamagnetic transition is absent in $M_c(H)$. These observations confirm the easy-plane AFM configuration of the Eu lattice. The saturated magnetization ($M_\mathrm{sat}$) for $M_c(H)$ is about 14.2 $\mu_B$/f.u., consistent with the presence of two Eu$^{2+}$ ions in the chemical formula. However, $M_\mathrm{sat}$ for $M_{ab}(H)$ is only about 11.7 $\mu_B$/f.u. This inconsistency is due to the mass uncertainty of the crystals used in the measurement, as the mass of a typical \EMCP\ crystal is less than 0.1 mg. Notably, the Curie--Weiss analysis of the polycrystalline sample, combined with its $M(H)$ at 1.8 K, yields consistent values for both $\mu_\mathrm{eff}$ and $M_\mathrm{sat}$ compared to those obtained from the analysis of the $c$-axis single-crystal data (provided in Fig. S4 of the SM~\cite{suppmatt}).

\subsection{Heat Capacity}

\begin{figure}
	\includegraphics[width=0.5\textwidth]{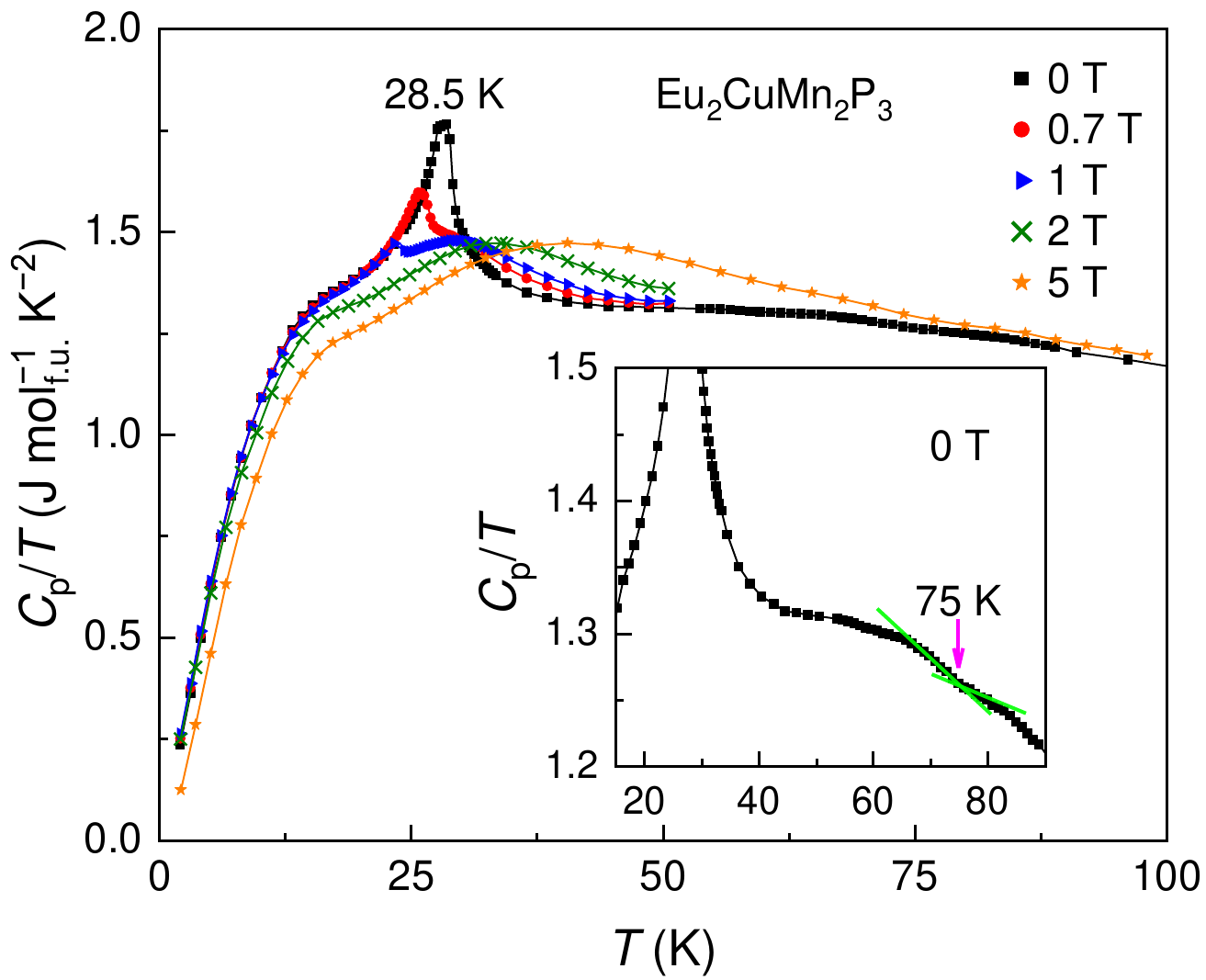}
	\caption{\label{fig:HeatCapacity}
			Specific heat divided by temperature ($C_\mathrm{p}/T$) as a function of temperature for the \EMCP\ single crystal from 2 to 100 K. The inset provides an enlarged view of the 0 T data, specifically highlighting the slope inflection near 75 K (indicated by the intersection of green guide lines) and the broad hump feature at lower temperatures.}
			
\end{figure}

To verify the magnetic transitions in \EMCP, as indicated by the magnetization data, we measured the specific heat $C_\mathrm{p}$ and plot $C_\mathrm{p}/T$ versus $T$ in Fig. \ref{fig:HeatCapacity}. The peak at 28.5 K aligns with the magnetic transition at $T_\mathrm{N}^\mathrm{Eu}$, confirming it as a bulk AFM transition. Upon applying fields, the peak diminishes and virtually disappears at 2 T, resulting in a weight shift of heat capacity from lower to higher temperatures. Unlike the sharp peak at 28.5 K, no heat capacity anomaly corresponding to the kink in $\chi_{ab}$ at $T_\mathrm{SR}$ is observed, supporting the notion that this transition arises from in-plane spin reorientation. The broad shoulder around 12 K is typical in $S = 7/2$ magnets, and can be explained using molecular field theory~\cite{johnston2015}. The inset reveals a broad hump in the zero-field data, arising from phonon contributions, along with a subtle slope inflection at $\sim$75 K (marked by the intersection of green guide lines). These features, the slope change and moderate rise below 75 K, are subtle yet reproducible, and cannot be explained by experimental artifacts (e.g., temperature step changes) or purely phononic effects. Instead, they may correlate with the AFM Mn ordering at $T_\mathrm{N}^\mathrm{Mn}$. The lack of a distinct transition peak is likely due to the quasi-2D Mn-Mn magnetic correlations discussed earlier, which do not generate a significant heat capacity signal~\cite{T_N,T_N_2}. Because of the limited mass of the \EMCP\ crystals used for heat capacity measurements, the data in the high-temperature region are of lower quality, making a precise analysis of magnetic entropy unavailable.

\subsection{Magnetic Energies}

\begin{figure*}
	\includegraphics[width=0.8\textwidth]{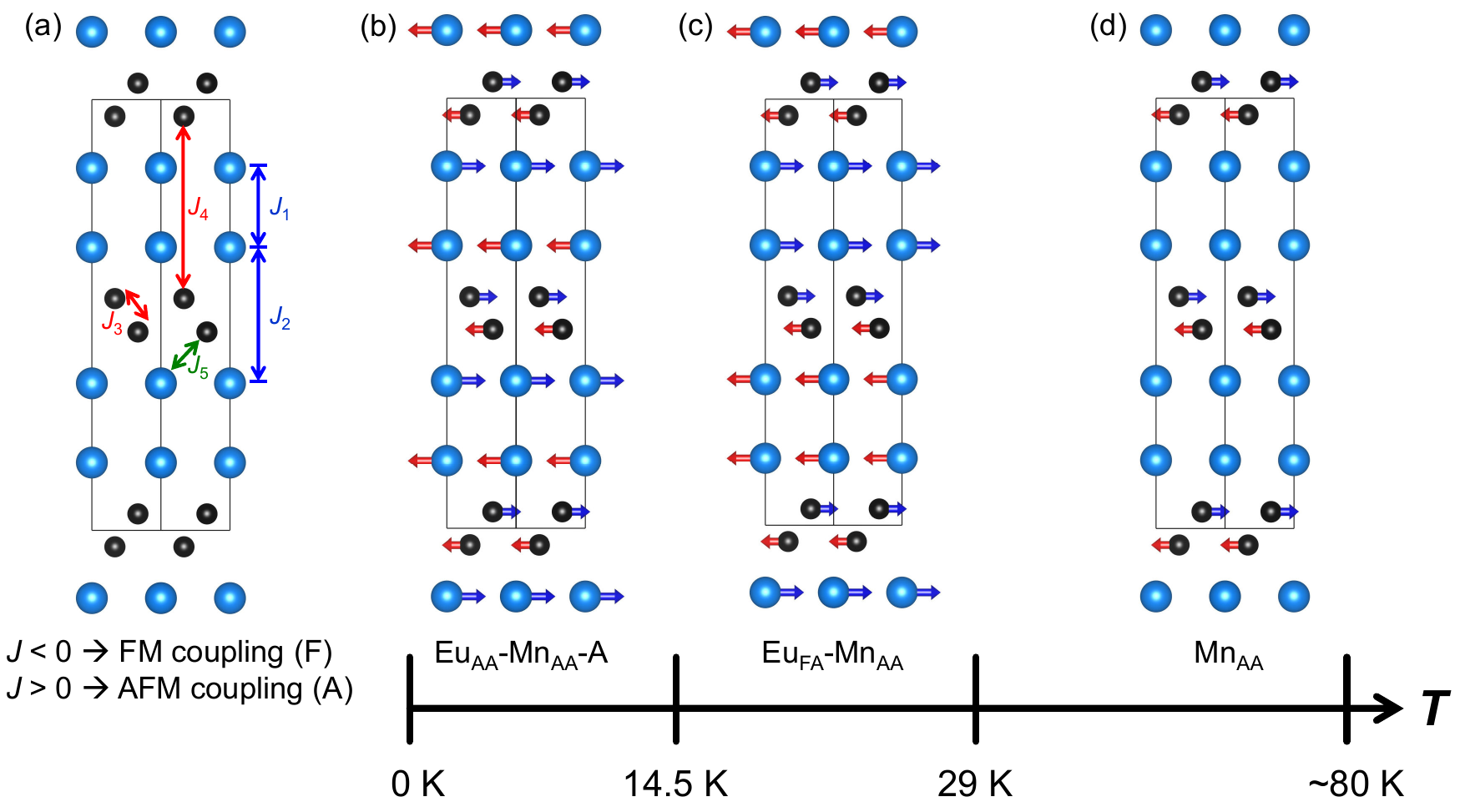}
	\caption{\label{fig:SPIN} Proposed AFM spin configurations in different temperature regions. (a) Some important inter-plane magnetic correlations. Negative and positive correlations represent FM and AFM couplings, respectively, whereas the dominant in-plane Eu-Eu and Mn-Mn FM couplings are not shown. (b) Magnetic structure of \EMCP\ below 14.5 K. (c) A possible magnetic structure between 14.5 K and 29 K. (d) AFM coupling between the Mn planes below 80 K.}
\end{figure*}

To gain deeper insights into the magnetic structure of \EMCP, we performed first-principles calculations to determine the magnetic energies for various spin configurations. Five representative inter-plane interactions are depicted in Fig. \ref{fig:SPIN}(a). Previous studies have shown that the Eu spins in both \ECP\ and \EMP\ are collinear~\cite{iha2019,wang2023,May2023,berry2023,yuan2024}, and the Mn spins in \ch{SrMn2As2} and \ch{EuMn2As2} also exhibit collinearity~\cite{dahal2019,das2016}. Therefore, we evaluated the magnetic energies for 15 collinear spin structures by combining different magnetic interactions, as summarized in Table \ref{tbl:energy}.  The lowest magnetic energy in the table is set to zero. It is noted that treating the Mn sites as nonmagnetic in combination with Eu$_\mathrm{AA}$ results in a significantly higher energy (7.15 eV/u.c.) compared to the Eu$_\mathrm{AA}$-Mn$_\mathrm{AA}$-A state listed in Table \ref{tbl:energy}. Additionally, attempts to combine nonmagnetic Mn with Eu$_\mathrm{FA}$ and Eu$_\mathrm{AF}$ configurations fail to converge. This observation supports our assertion that strong AFM correlations are present at the Mn sites. Varying the effective Hubbard correlation $U$ values for Mn $3d$ orbitals ($U_\mathrm{Mn}$) does not significantly affect our conclusions regarding the magnetic structure of \EMCP. The detailed results are provided in Fig. S7 of the SM~\cite{suppmatt}. Moreover, the energies were calculated without considering spin--orbit coupling (SOC), as its inclusion has a negligible effect on the relative magnitudes of the magnetic energies.

\begin{table*}
	\caption{Calculated magnetic energies per unit cell (meV/u.c.) for \EMCP\ with $U_\mathrm{Mn}$ = 2.0 eV, combining Eu and Mn sublattices in various spin configurations as defined by interactions illustrated in Fig. \ref{fig:SPIN}(a).}
		
	\begin{ruledtabular}\label{tbl:energy}
	\begin{tabular}{ccccc}
		& Mn$_\mathrm{FF}$\footnote{Mn$_{J_3J_4}$ denotes the couplings between Mn planes within the \EMP\ BL and across the \ECP\ BL. For instance, Mn$_\mathrm{FF}$ means $J_3<0$ and $J_4<0$.} &Mn$_\mathrm{AA} $ & Mn$_\mathrm{FA}$ & Mn$_\mathrm{AF}$  \\
		\hline
		Eu$_\mathrm{AA}$\footnote{Eu$_{J_1J_2}$ signifies the interlayer Eu-Eu couplings within the \ECP\ and \EMP\ BLs. For instance, Eu$_\mathrm{AA}$ means $J_1>0$ and $J_2>0$.} & 1200.2  & 0 (A), 17.2 (F)\footnote{In some cells, two values are provided to differentiate the couplings between the nearest Eu and Mn layers (A: $J_5>0$, F: $J_5<0$).} & 1197.2 & 6.2\\
		Eu$_\mathrm{FA}$ & 1197.3  & 2.1 & 1192.8 & 3.4 (A), 8.0 (F)  \\
		Eu$_\mathrm{AF}$ & 1205.5  & 4.2 & 1237.2 (A), 1169.2 (F) & 3.6 \\
	\end{tabular}	
	\end{ruledtabular}

\end{table*}

The resulting energy values are listed in Table \ref{tbl:energy}, with the nomenclature of the spin structures explained in the footnotes. The energy values in the Mn$_\mathrm{AA}$ and Mn$_\mathrm{AF}$ columns, both characterized by $J_3>0$, are substantially lower than those in the Mn$_\mathrm{FF}$ and Mn$_\mathrm{FA}$ columns. This suggests strong AFM coupling between the nearest Mn planes, similar to the inter-plane Mn-Mn coupling reported in \EMP\ and \ch{EuMn2As2}~\cite{berry2023,dahal2019}. Additionally, the small energy difference between the Mn$_\mathrm{AA}$ and Mn$_\mathrm{AF}$ columns indicates that the Mn-Mn coupling across the \ECP\ BL ($J_4$) is far weaker compared to that within the \EMP\ BL ($J_3$), which is not surprising given the large inter-B-L Mn-Mn distances. This observation supports our argument for the quasi-2D Mn correlations. Given that the lowest energy is found in the Mn$_\mathrm{AA}$ column and considering the Mn-Mn couplings in \EMP\ and \ch{EuMn2As2}, the Mn$_\mathrm{AA}$ configuration likely reflects the true Mn spin structure in \EMCP. Within the Mn$_\mathrm{AA}$ column, the Eu$_\mathrm{AA}$-Mn$_\mathrm{AA}$-A combination exhibits the lowest energy, while the Eu$_\mathrm{FA}$-Mn$_\mathrm{AA}$ combination shows a slightly higher energy. The close energy values suggest strong competition among different magnetic phases. Since magnetocrystalline anisotropy is temperature-dependent, its competition with exchange interactions can cause the spin structure to transition between phases as the temperature varies, as shown in Fig. \ref{fig:SPIN}. As the temperature decreases, AFM ordering at the Mn sites is initially established at $T_\mathrm{N}^\mathrm{Mn}$ = 80 K. The $T_\mathrm{C}$ of \ECP\ is 32 K, roughly twice the $T_\mathrm{N}$ of 17 K for \EMP. Given the proximity of the transition temperatures, it is reasonable to associate the AFM transition at 29 K with the \ECP\ BL and the spin-reorientation transition at 14.5 K with the \EMP\ BL. Thus, the transition at 29 K, though overall AFM, may involve ferromagnetically coupled Eu planes ($J_1<0$), specifically the Eu$_\mathrm{FA}$-Mn$_\mathrm{AA}$ spin structure, which is driven by the inter-plane Eu-Eu FM correlations within \ECP\ BLs. As the temperature decreases further, the evolving magnetocrystalline anisotropy may increasingly compete with exchange interactions, ultimately leading to a spin-reorientation transition to the Eu$_\mathrm{AA}$-Mn$_\mathrm{AA}$-A structure at 14.5 K. 

It is worth noting that the spin structure Eu$_\mathrm{FA}$-Mn$_\mathrm{AF}$-A also exhibits an energy comparable to that of Eu$_\mathrm{FA}$-Mn$_\mathrm{AA}$. Despite this, we have opted not to consider it as a prime candidate, given the absence of a spin-reorientation transition between Mn$_\mathrm{AF}$ and the magnetic ground state Mn$_\mathrm{AA}$. Nonetheless, it remains crucial to acknowledge that while the proposed spin structures and their evolution, inferred from magnetic energy calculations and experimental observations, provide a plausible explanation, they necessitate further validation using methods such as neutron diffraction.

\subsection{Electrical Resistivity}

\begin{figure*}
	\includegraphics[width=0.75\textwidth]{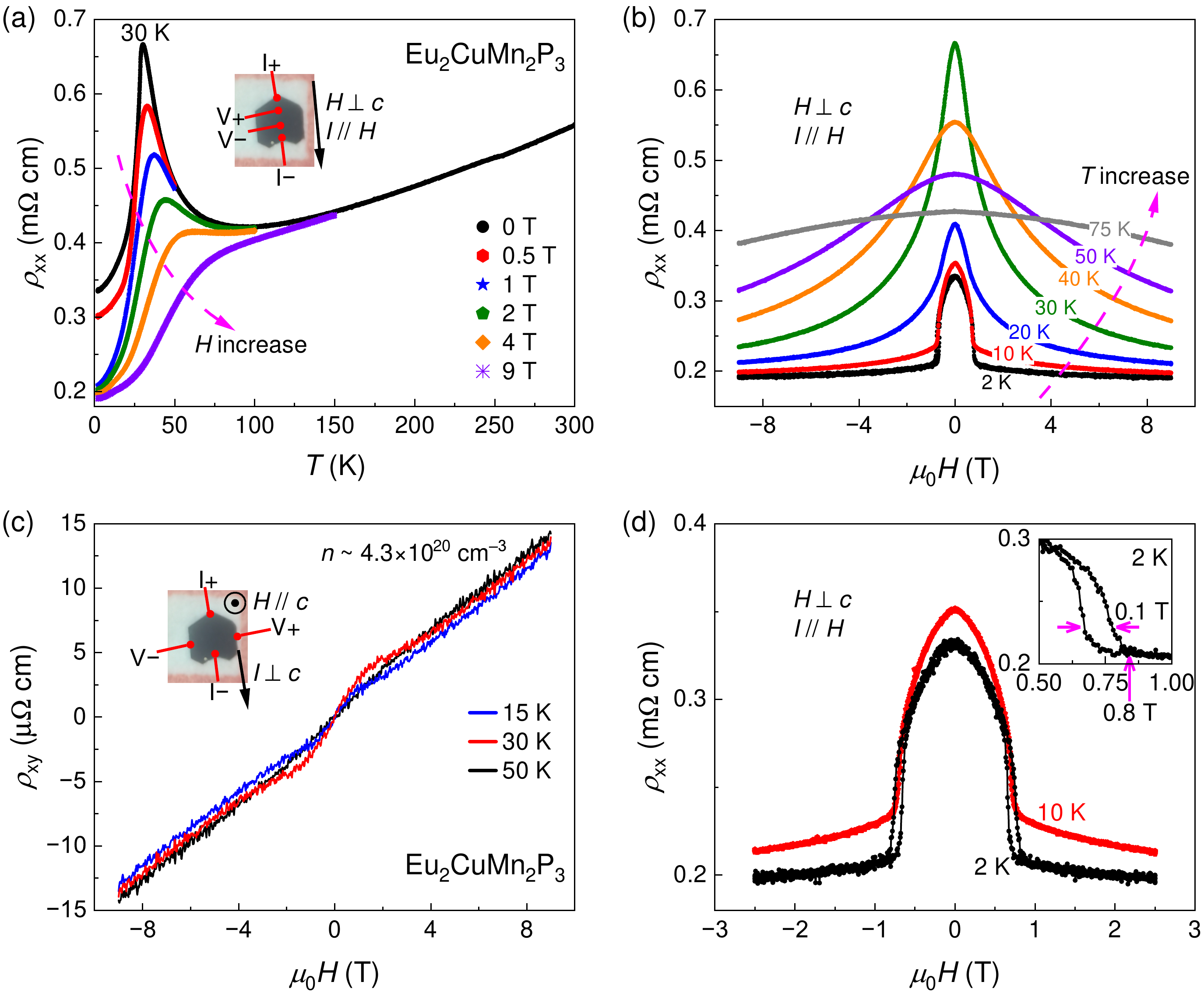}
	\caption{\label{fig:RESISTIVITY} (a) Temperature-dependent in-plane resistivity under various applied magnetic fields. The inset shows a photo of the \EMCP\ crystal and the electrode configuration. (b) Field-dependent resistivity curves from 2 to 75 K. (c) Magnetic field dependence of the Hall resistivity of \EMCP\ at 15 K, 30 K, and 50 K. (d) Hysteresis in resistivity at low temperatures.}
\end{figure*}

Previous studies have shown that \EMP\ is an insulator with a band gap of 0.2 eV~\cite{berry2023}, while \ECP\ exhibits a metallic state with a hole carrier concentration of $\sim10^{20}$ cm$^{-3}$~\cite{wang2023,yuan2024}. As an intergrowth of \ECP\ and \EMP, \EMCP\ shares similar transport properties with \ECP\ but also displays unique characteristics. Figure \ref{fig:RESISTIVITY}(a) illustrates the temperature dependence of the longitudinal resistivity $\rho_{xx}(T)$ under in-plane magnetic fields ($H\perp c$). At zero field, $\rho_{xx}(T)$ decreases gradually above 100 K and then shows a prominent peak at 30 K, consistent with the AFM transition at $T_\mathrm{N}^\mathrm{Eu}$. The pronounced increase in resistivity above 30 K is attributed to enhanced magnetic fluctuations before the onset of magnetic order. Below $T_\mathrm{N}^\mathrm{Eu}$, $\rho_{xx}(T)$ drops sharply due to reduced magnetic scattering. This distinctive peak behavior, indicative of magnetic ordering, is also observed in \ECP\ and other Eu-based materials with low carrier densities~\cite{wang2023,zhou2024,chen2024,yuan2024}. At $T_\mathrm{N}^\mathrm{Mn}$ of 80 K and $T_\mathrm{SR}$ of 14.5 K, no noticeable anomalies are observed, suggesting that the AFM ordering of Mn sublattice and the spin-reorientation transition of Eu sublattice have minimal effects on magnetic scattering. When a magnetic field is applied, the resistivity peak is progressively suppressed, demonstrating a notable negative magnetoresistance (nMR) effect. The field-dependent resistivity $\rho_{xx}(H)$ curves from 2 to 75 K are shown in Fig. \ref{fig:RESISTIVITY}(b). The nMR effect is insignificant at 75 K, but becomes more pronounced as the temperature decreases. At 30 K, the nMR effect reaches its maximum, with $\rho_{xx}$ decreasing by 65\% under a field of 9 T. Below $T_\mathrm{N}^\mathrm{Eu}$, the nMR effect diminishes gradually due to reduced spin-dependent scattering. The data of $\rho_{xx}(T)$ and $\rho_{xx}(H)$ with out-of-plane fields ($H\parallel c$) are presented in Fig. S5 of the SM~\cite{suppmatt}, which only shows a small anisotropy from the data in Fig. \ref{fig:RESISTIVITY}.

The low-temperature $\rho_{xx}(H)$ curves of \EMCP\ display distinct features compared to those of \ECP\ and other Eu-based layered materials. First, the $\rho_{xx}(H)$ curves at 10 K and 2 K resemble a pointed arch, characterized by a gradual decline at lower fields followed by a sharp drop at higher fields, as shown in Fig. \ref{fig:RESISTIVITY}(d). This behavior contrasts with the Lorentzian-like peaks observed in $\rho_{xx}(H)$ curves above 20 K. The unusual low-temperature behavior of $\rho_{xx}(H)$ may be related to the altered in-plane spin alignment below the spin-reorientation transition at 14.5 K, leading to a different field response. A corroboration of this is that the $\rho_{xx}(H)$ curve at 2 K with $H\parallel c$, shown in Fig. S5(c), does not exhibit an arch-shaped dependence. Second, a resistivity hysteresis is evident in the 2 K and 10 K curves within the field range of 0.5 to 0.8 T. This field range aligns well with the reopened magnetic hysteresis loop seen in Fig. \ref{fig:SUSCEPTIBILITY}(e), indicating that this phenomenon is directly linked to the metamagnetic transition and the FM component. The maximum gap in the resistivity hysteresis is about 0.1 T, as illustrated in the inset of Fig. \ref{fig:RESISTIVITY}(d). In contrast, for $\rho_{xx}(H)$ with $H\parallel c$ (Fig. S5(c)), the resistivity hysteresis effect is much reduced, consistent with the negligible magnetic hysteresis near the $c$-axis saturation field shown in Fig. \ref{fig:SUSCEPTIBILITY}(f). We also measured the $\rho_{xx}(T)$ and $\rho_{xx}(H)$ data for the polycrystalline samples of \EMCP, presented in Fig. S6 of the SM~\cite{suppmatt}, which are in line with those observed in single crystals.

The field dependence of the Hall resistivity $\rho_{xy}(H)$ at 15 K, 30 K, and 50 K for \EMCP\ is shown in Fig. \ref{fig:RESISTIVITY}(c), with in-plane current and out-of-plane fields. At high magnetic fields, such as above 4 T, the $\rho_{xy}(H)$ curves are primarily governed by the ordinary Hall effect (OHE), displaying a linear relationship. The three curves exhibit a consistent positive slope of $\sim1.5$ $\mu\Omega$ cm T$^{-1}$ at high fields, indicating a nearly constant carrier concentration dominated by holes, with a value of $4.3\times10^{20}$ cm$^{-3}$. This carrier concentration is similar to that observed in \ECP, and we note that the room-temperature resistivity of \EMCP, 0.55 m$\Omega$ cm, closely matches that of \ECP~\cite{wang2023,yuan2024}. Given that \EMP\ is insulating, these observations suggest that the \ECP\ BL serves as the primary source of carriers in \EMCP. Furthermore, like \ECP, the anomalous Hall effect (AHE) contributes to the $\rho_{xy}(H)$ curves at low fields, reaching its maximum around the magnetic ordering temperature at $T_\mathrm{N}^\mathrm{Eu}$~\cite{wang2023,yuan2024}.

\subsection{Electronic Structure}

\begin{figure*}
	\includegraphics[width=0.75\textwidth]{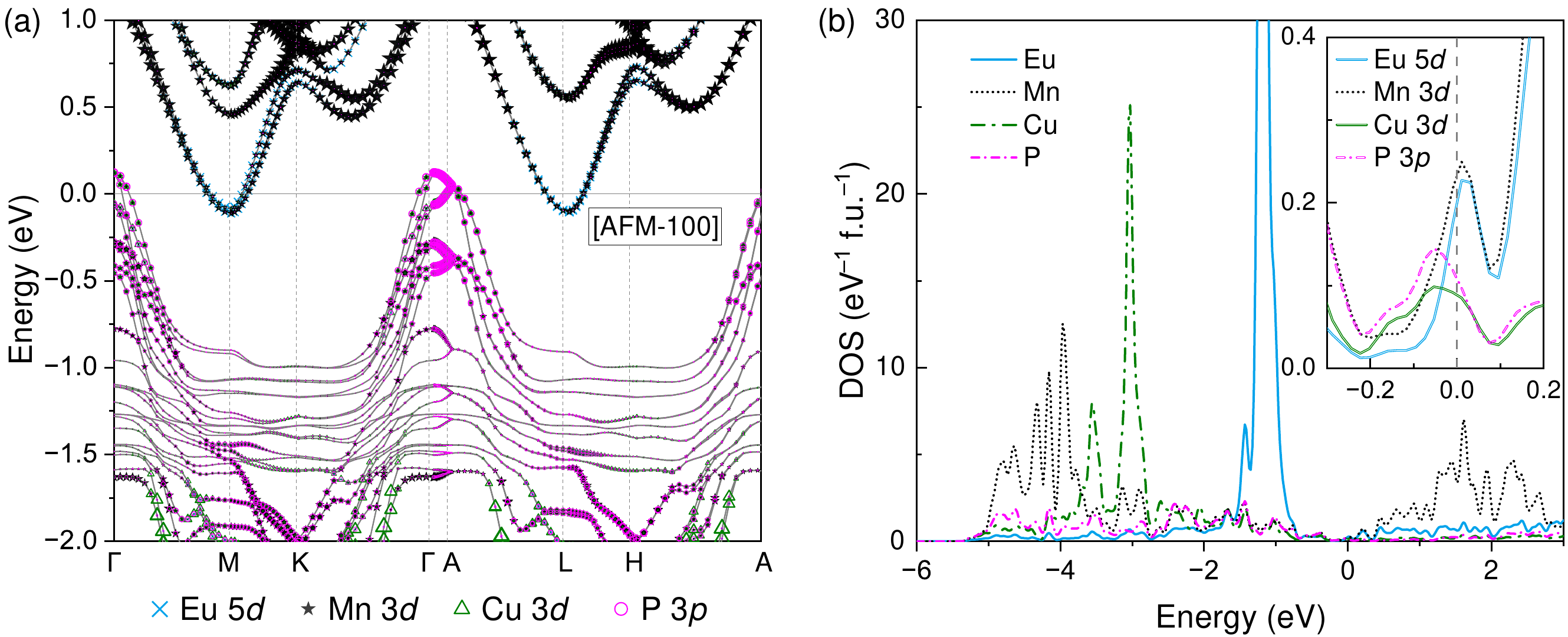}
	\caption{\label{fig:DFT}
		(a) Calculated band structure of \EMCP\ with SOC. (b) DOS for \EMCP. The inset zooms in on the DOS of selected orbitals around the Fermi energy.}
\end{figure*}

To better understand electronic properties of \EMCP, we calculated its band structure using the Eu$_\mathrm{AA}$-Mn$_\mathrm{AA}$-A spin configuration, which represents the magnetic ground state. We examined the variations in the band structure by considering different spin orientations and Hubbard $U_\mathrm{Mn}$ values. While the spin orientation has a negligible effect on the band structure, the $U_\mathrm{Mn}$ value significantly influences it. The results are summarized in Fig. S8 of the SM~\cite{suppmatt}. Calculations with $U_\mathrm{Eu}$ = 5.0 eV and $U_\mathrm{Mn}$ = 5.0 eV predict a narrow band gap of 50--60 meV, which contradicts the observed metallic behavior. However, setting $U_\mathrm{Eu}$ = 5.0 eV and $U_\mathrm{Mn}$ = 2.0 eV yields a semimetallic band structure that aligns well with experimental observations. Therefore, a moderate $U_\mathrm{Mn}$ value better describes the electron correlation of the Mn $3d$ orbitals. Additionally, the corresponding Mn magnetic moment is 4.1 $\mu_\mathrm{B}$/Mn, consistent with the local moment picture for Mn.

The calculated band structure and density of states (DOS) with $U_\mathrm{Mn}$ = 2.0 eV are presented in Fig. \ref{fig:DFT}. The resulting band structure reveals that both electron and hole pockets contribute to electrical conduction, although the Hall transport behavior is predominantly influenced by hole carriers. The electron pockets are chiefly formed by the Eu $5d$ and Mn $3d$ orbitals, while the hole pockets are primarily composed of the Cu $3d$ and P $3p$ orbitals, suggesting an electron transfer from the \ECP\ BL to \EMP\ BL. The Eu $4f$ states are predominantly located between 1.0 and 1.5 eV below the Fermi level ($E_\mathrm{F}$), as depicted in Fig. \ref{fig:DFT}(b), and contribute minimally to the states near the $E_\mathrm{F}$. At $E_\mathrm{F}$, \EMCP\ exhibits a small DOS of 1.3 states eV$^{-1}$ f.u.$^{-1}$, consistent with the low carrier density revealed by the Hall resistivity measurements. An enlarged view of the DOS near $E_\mathrm{F}$, shown in the inset of Fig. \ref{fig:DFT}(b), reveals evident hybridization between the Eu $5d$, Mn $3d$, Cu $3d$, and P $3p$ orbitals. This hybridization accounts for the reduced magnetic moment of 4.1 $\mu_\mathrm{B}$/Mn in our calculation and the metallic nature of \EMCP.

\section{Conclusions}

In summary, we report the design and synthesis of a novel layered intergrowth compound, \EMCP, achieved by alternately stacking the fundamental structural units of \ECP\ and \EMP. This structural hybridization not only retains the inherent magnetic correlations within the \ECP\ and \EMP\ BLs but also introduces new inter-B-L magnetic interactions, leading to complex and intriguing magnetic behaviors in \EMCP. Our magnetization studies (both dc and ac) reveal distinct magnetic transitions: AFM ordering of Mn at 80 K ($T_\mathrm{N}^\mathrm{Mn}$), AFM ordering of Eu at 29 K ($T_\mathrm{N}^\mathrm{Eu}$), and a spin-reorientation transition within the $ab$ plane at 14.5 K ($T_\mathrm{SR}$). Notably, weak ferromagnetism associated with Mn is observed below $T_\mathrm{N}^\mathrm{Mn}$, possibly arising from spin canting or itinerant carriers. Hysteresis loops in the $M(H)$ curves are evident at low temperatures, with a unique reopening of the loop when $H\perp c$ after a metamagnetic transition at 0.6 T. These magnetic transitions at $T_\mathrm{N}^\mathrm{Mn}$ and $T_\mathrm{N}^\mathrm{Eu}$ are corroborated by heat capacity measurements. The subtle peaks around $T_\mathrm{N}^\mathrm{Mn}$ in susceptibility and heat capacity data suggest that the presence of Mn ordering and Mn-Mn correlations are quasi-2D, possibly due to the large Mn-Mn separation across the \ECP\ BL. To explore the spin structure of \EMCP, we conducted calculations of magnetic energies for various collinear configurations. Our findings indicate that the ground state involves AFM couplings between inter-plane Eu-Eu, Mn-Mn, and Eu-Mn (Eu$_\mathrm{AA}$-Mn$_\mathrm{AA}$-A), while the spin configurations Mn$_\mathrm{AA}$ and Eu$_\mathrm{FA}$-Mn$_\mathrm{AA}$ can account for the spin structures below $T_\mathrm{N}^\mathrm{Mn}$ and between $T_\mathrm{N}^\mathrm{Eu}$ and $T_\mathrm{SR}$, respectively. The transport properties of \EMCP\ resemble those of \ECP, characterized by a pronounced nMR effect and a similar carrier density dominated by holes. A notable distinction is the arch-like $\rho_{xx}(H)$ dependence at low temperatures and the resistivity hysteresis, both of which are linked to the spin configuration below $T_\mathrm{SR}$ and the metamagnetic transition. Lastly, we computed the band structure of \EMCP\ using the Eu$_\mathrm{AA}$-Mn$_\mathrm{AA}$-A spin configuration. The resulting semimetallic band structure aligns well with experimental observations, revealing that both electron and hole pockets contribute to electrical conduction. Additionally, exploring the temperature-dependent evolution of \EMCP's magnetic structure using techniques such as neutron diffraction and resonant elastic x-ray scattering (REXS) will be both interesting and crucial.

The structural design of \EMCP\ adheres to the principles of our B-L model, which integrates material BLs with closely matched in-plane lattice parameters and similar crystal symmetries. This strategy minimizes deformation within the constituent BLs. The successful synthesis of \EMCP\ underscores the practicality and efficacy of the B-L model for exploring novel intergrowth compounds. It also paves the way for discovering a wide range of materials isostructural to \EMCP, leveraging the high compatibility of widely available SrPtSb-type and \ch{CaAl2Si2}-type compounds~\cite{klufers1977,mewis1978,klufers1980,mewis1978a,tomuschat1981}. Beyond this, the assembly of BLs offers an effective means to introduce novel magnetic correlations or to tune existing ones. In the case of \EMCP, Mn ordering is observed, although whether it is long-range or short-range remains undetermined. This is primarily due to altered inter-B-L Mn-Mn coupling and modified chemical pressure within the intergrowth structure, as compared to \EMP~\cite{berry2023}. Looking forward, we anticipate that the B-L model will facilitate the discovery of more intergrowth compounds with unique magnetic behaviors through strategic B-L combinations. This approach promises to enrich the field of materials science, offering fresh opportunities for both fundamental research and practical applications.

\textit{Note added.} During the peer review process, we became aware of recent work characterizing \ECZP\ (a stacking of \ECP\ and \EZP)~\cite{may2025}, which is isostructural to \EMCP\ and could serve as an excellent complement to our study. \ECZP\ exhibits an AFM transition at 40.3~K, attributed to ordering of the Eu sublattice. The lack of FM signatures in \ECZP\ corroborates our conclusion that the FM behavior in \EMCP\ originates exclusively from the Mn sublattice. A weak anomaly is observed at $\sim$26~K (near $T_{\mathrm{N}}$ of \EZP) in the in-plane susceptibility ($H \perp c$) of \ECZP\ under a 25~Oe field, reminiscent of the spin reorientation transition in \EMCP\ at 14.5~K. Furthermore, neutron diffraction reveals that \ECZP\ adopts an up-up-down-down AFM spin structure, equivalent to the Eu$_{\mathrm{FA}}$ structure discussed in this work. This finding supports the preservation of parent compound spin couplings in the intergrowth structure.

\begin{acknowledgments}
This work was supported by the National Natural Science Foundation of China (Grants No. 12204094, No. 12325401, and No. 123B2053), the Natural Science Foundation of Jiangsu Province (Grant No. BK20220796), the Start-up Research Fund of Southeast University (Grant No. RF1028623289), the SEU Innovation Capability Enhancement Plan for Doctoral Students (Grant No. CXJH\_SEU 25136), and the Big Data Computing Center of Southeast University.

\end{acknowledgments}

%
%
%
%
\bibliography{references}

\end{document}


\title{\underline{Supplemental Materials:}\\ Structural design and multiple magnetic orderings of the intergrowth compound \EMCP}

\author{Xiyu Chen}
\altaffiliation{The authors contributed equally to this work.}
\affiliation{Key Laboratory of Quantum Materials and Devices of Ministry of Education, School of Physics, Southeast University, Nanjing 211189, China}
\author{Ziwen Wang}
\altaffiliation{The authors contributed equally to this work.}
\affiliation{Key Laboratory of Quantum Materials and Devices of Ministry of Education, School of Physics, Southeast University, Nanjing 211189, China}
\author{Wuzhang Yang}
\affiliation{School of Science, Westlake University, Hangzhou 310024, China}
\affiliation{Institute of Natural Sciences, Westlake Institute for Advanced Study, Hangzhou 310024, China}
\author{Jia-Yi Lu}
\affiliation{School of Physics, Interdisciplinary Center for Quantum Information and State Key Laboratory of Silicon and Advanced Semiconductor Materials, Zhejiang University, Hangzhou 310058, China}
\author{Zhiyu Zhou}
\affiliation{Key Laboratory of Quantum Materials and Devices of Ministry of Education, School of Physics, Southeast University, Nanjing 211189, China}
\author{Zhi Ren}
\affiliation{School of Science, Westlake University, Hangzhou 310024, China}
\affiliation{Institute of Natural Sciences, Westlake Institute for Advanced Study, Hangzhou 310024, China}
\author{Guang-Han Cao}
\affiliation{School of Physics, Interdisciplinary Center for Quantum Information and State Key Laboratory of Silicon and Advanced Semiconductor Materials, Zhejiang University, Hangzhou 310058, China}
\affiliation{Collaborative Innovation Center of Advanced Microstructures, Nanjing University, Nanjing 210093, China}
\author{Shuai Dong}
\email{sdong@seu.edu.cn}
\affiliation{Key Laboratory of Quantum Materials and Devices of Ministry of Education, School of Physics, Southeast University, Nanjing 211189, China}
\author{Zhi-Cheng Wang}
\email{wzc@seu.edu.cn}
\affiliation{Key Laboratory of Quantum Materials and Devices of Ministry of Education, School of Physics, Southeast University, Nanjing 211189, China}
\date{\today}

\maketitle

\clearpage

\begin{figure}[!htb]
	\includegraphics[width=1.0\textwidth]{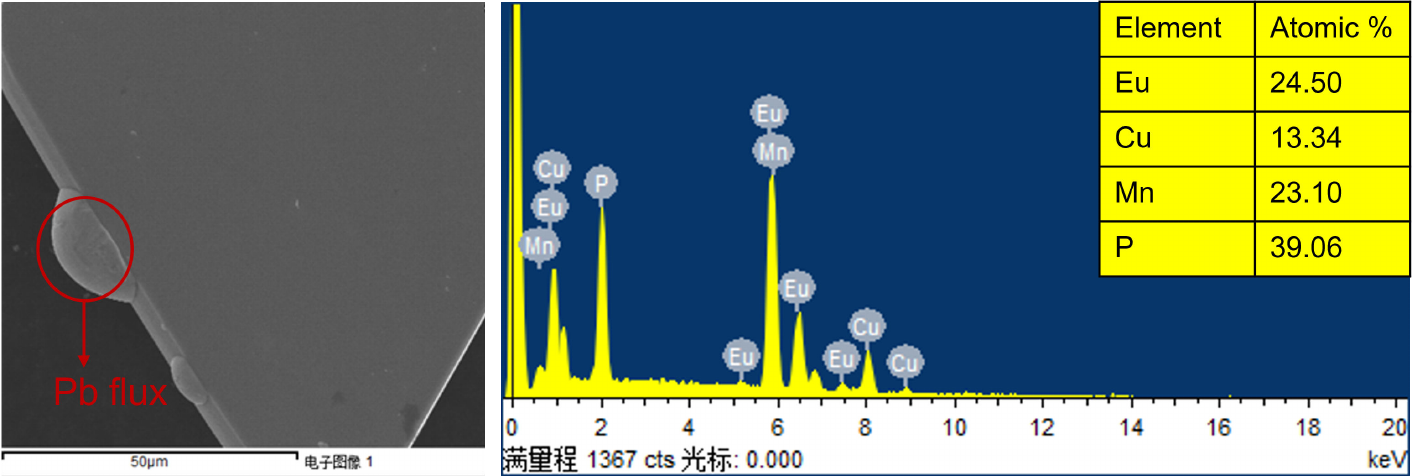}
	\caption{\label{fig:EDX}
		A representative scanning electron microscopy (SEM) image and corresponding energy-dispersive x-ray spectroscopy (EDX) spectrum of a \EMCP\ single crystal.
	}
\end{figure}

\begin{center}
	\textbf{Contents}
	
	\begin{enumerate}
		\item Figure \ref{fig:EDX} and Table \ref{TabS1}: Chemical composition analysis of \EMCP
		\item Figure \ref{fig:CW_analysis}: Supplementary Curie--Weiss analysis and $M(H)$ curves
		\item Figure \ref{fig:PXRD}: Powder x-ray diffraction pattern for \EMCP\ at room temperature
		\item Figure \ref{fig:PCmagnetization}: Curie--Weiss analysis and $M(H)$ curves for the polycrystalline \EMCP	
		\item Figures \ref{fig:SX_c_Resistivity} and \ref{fig:PResistivity}: Supplementary resistivity data for single crystal with $c$-axis fields and polycrystalline sample
		\item Table \ref{tbl:1} and Figure \ref{fig:energies}: Cell parameters and magnetic energies calculated with different $U_\mathrm{Mn}$ values		
		\item Figure \ref{fig:band}: Band structures calculated with different $U_\mathrm{Mn}$ values and spin configurations
	\end{enumerate}
\end{center}
\clearpage

\subsection{Chemical composition analysis of \EMCP}

\begin{table}[htbp]
	\caption{\label{TabS1} Chemical composition analysis of \EMCP\ single crystals via EDX. The first column shows the data points and corresponding specimens, for example, “S2, P1” represents the first point from sample 2. The number of Mn atoms in the average formula unit is normalized to 2, and other data are rescaled accordingly. The resulting chemical composition, with uncertainties representing one standard deviation, was determined to be Eu$_{1.98(7)}$Cu$_{1.07(5)}$Mn$_{2.00(7)}$P$_{3.04(11)}$.}
	
	\begin{tabular}{c|c|c|c|c}
		\hline\hline
		& \qquad Eu \qquad\qquad & \qquad Cu\qquad\qquad & \qquad Mn\qquad\qquad & \qquad P \qquad\qquad \\
		\hline
		S1&	26.14 &	14.20 &	24.39 &	35.27\\
		\hline
		S2, P1&	23.85 &	13.18 &	24.61 &	38.36\\
		S2, P2&	24.16 &	12.14 &	25.87 &	37.82\\
		\hline
		S3&	24.64 &	13.48 &	25.66 &	36.22\\
		\hline
		S4&	23.63 &	12.89 &	24.72 &	38.76\\
		\hline
		S5 & 24.50 & 	13.34 & 23.10 & 39.06\\
		\hline
		average	&24.5 (0.8) &	13.2 (0.6)&	24.7(0.9)&	37.6(1.4)\\
		\hline
		composition &	1.98(7) &	1.07(5)&	2.00(7)	& 3.04(11)\\
		\hline\hline
	\end{tabular}
	

\end{table}
\clearpage

\subsection{Supplementary magnetic data for the single-crystal \EMCP}

The Curie--Weiss analysis was performed for both $\chi_{ab}(T)$ and $\chi_c(T)$ at 0.1 T from 200 to 350 K. The resulting Weiss temperature of $\theta=39.4$ K and effective moment of $\mu_\mathrm{eff}=10.83$ $\mu_B$/f.u. in panel (a) are consistent with the results in panel (b). The discrepancies in the values are primarily attributed to the mass uncertainty of the crystals used, given that the mass of a typical \EMCP\ crystal is less than 0.1 mg. The $c$-axis magnetization as a function of field $M_{c}(H)$ is also shown in panel (c) for reference. The $M(H)$ curves at 60 K for both $H\perp c$ and $H\parallel c$ are plotted in panel (d), highlighting the magnetic hysteresis. The saturated moment for the ferromagnetic (FM) component is linearly extrapolated to $0.22~\mu_\mathrm{B}/\mathrm{f.u.}$ ($0.11~\mu_\mathrm{B}/\mathrm{Mn}$) using the $c$-axis $M(H)$ curve at 60~K, which agrees well with the value ($0.19~\mu_\mathrm{B}/\mathrm{f.u.}$, $0.09~\mu_\mathrm{B}/\mathrm{Mn}$) reported in our manuscript through susceptibility analysis.

 \begin{figure}[!htb]
 \includegraphics[width=0.7\textwidth]{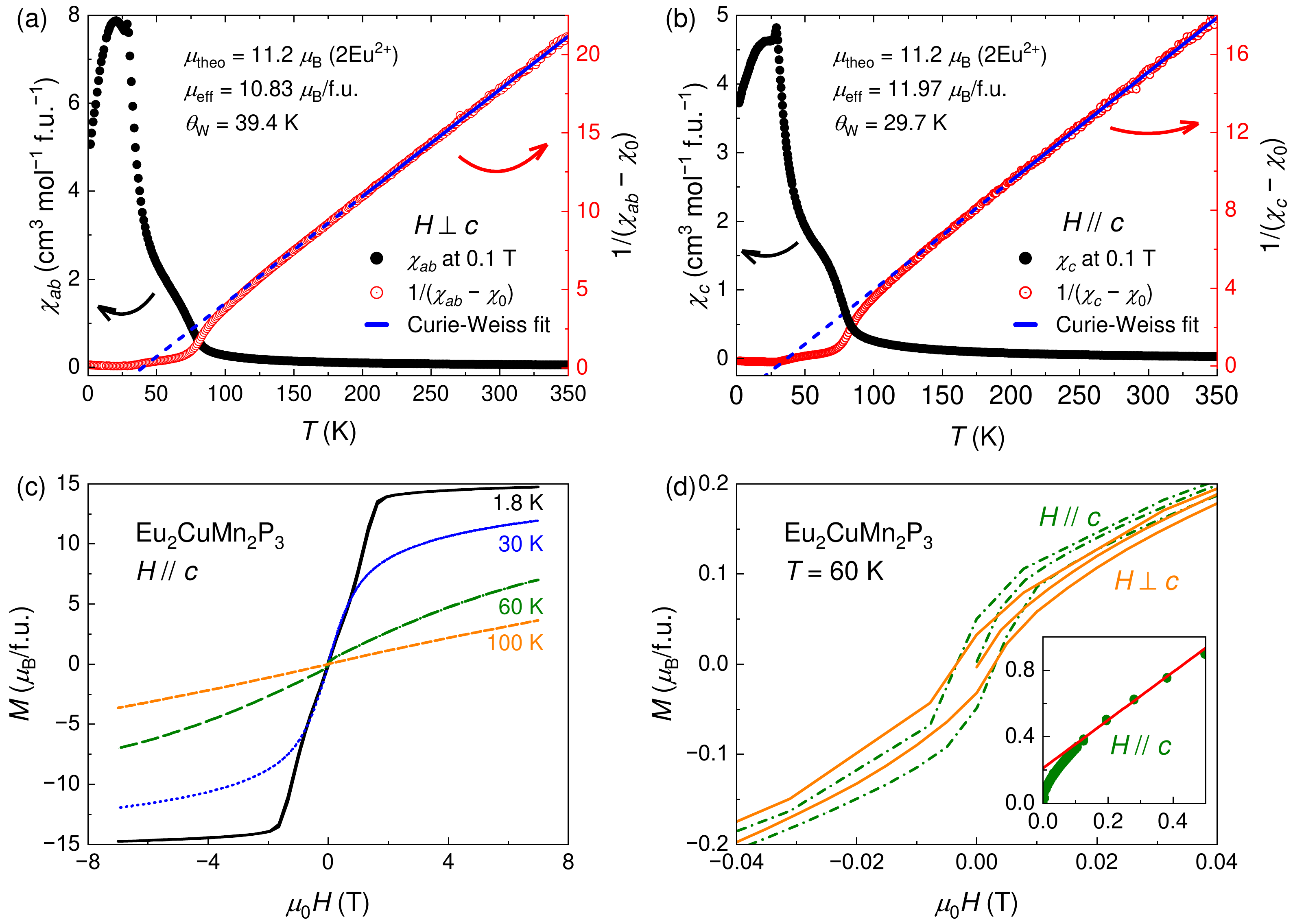}
 \caption{\label{fig:CW_analysis} 
 (a) In-plane susceptibility $\chi_{ab}$ (left axis) and the Curie--Weiss analysis (right axis) of \EMCP\ at 0.1 T. Black dots represent $\chi_{ab}$, while red circles show its inverse. The blue line is fit to $1/\chi_{ab}$. (b) Out-of-plane susceptibility $\chi_c$ (left axis) at 0.1 T and the corresponding Curie--Weiss analysis (right axis). (c) Out-of-plane magnetization $M_{c}$ as a function of magnetic field at several temperatures. (d) $M(H)$ curves with $H\perp c$ (orange) and $H\parallel c$ (olive) at 60 K. The inset displays a linear extrapolation of the $c$-axis $M(H)$ curve from above saturation to $H$ = 0, which allows extraction of the saturated magnetization for the FM component.
 }
 \end{figure}

\clearpage

\subsection{Powder x-ray diffraction pattern}
\begin{figure}[!htb]
	\includegraphics[width=0.8\textwidth]{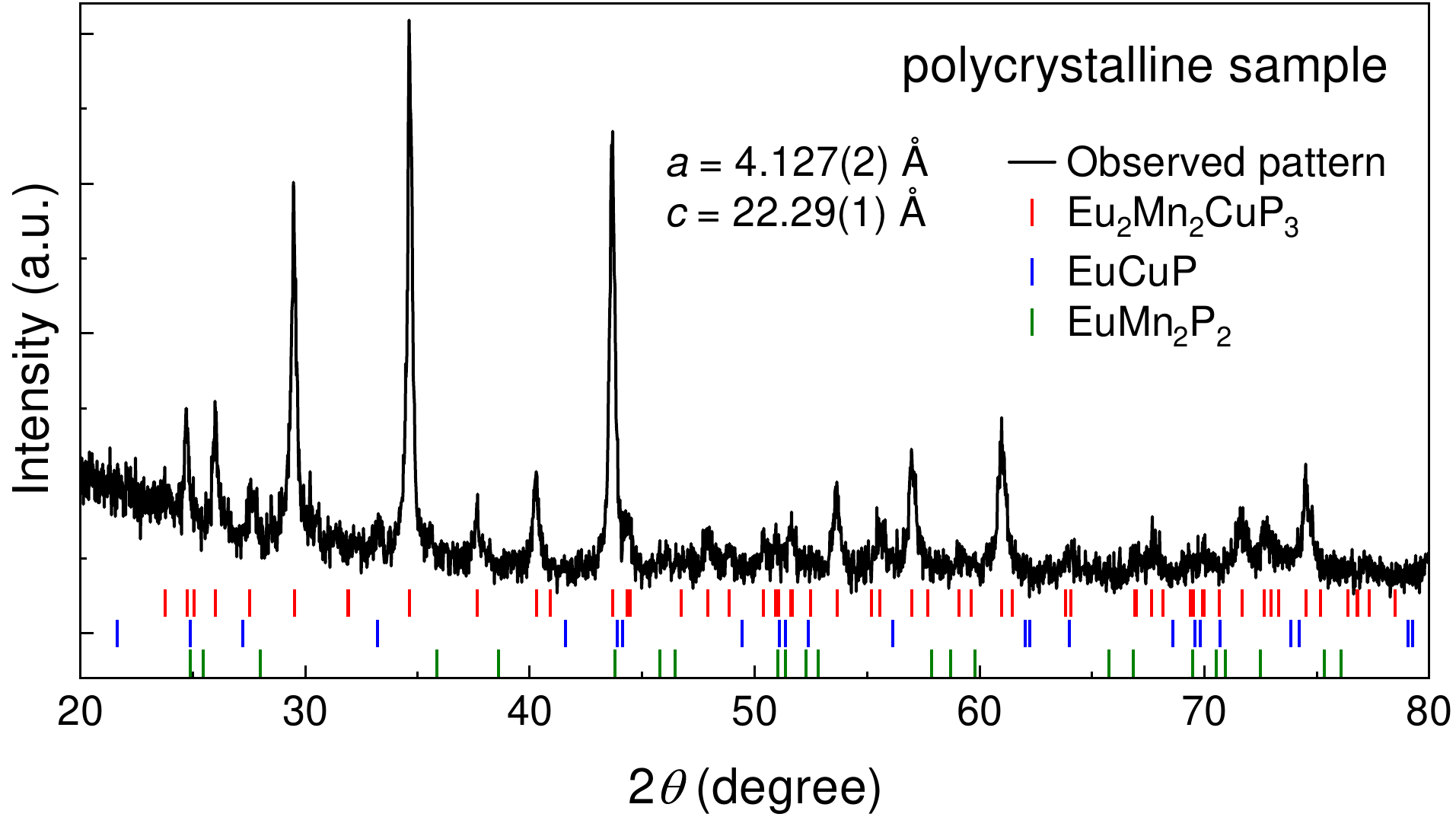}
	\caption{\label{fig:PXRD}
		Room-temperature powder XRD pattern (black curve) for \EMCP\ polycrystalline sample in the $2\theta$ range of $20^\circ$ to $80^\circ$. The $hkl$ indices for \EMCP\ (red), \ECP\ (blue) and \EMP\ (olive) are indicated by ticks below the pattern.
	}
\end{figure}

The powder x-ray diffraction (XRD) pattern for the \EMCP\ polycrystalline sample is shown in Fig.~\ref{fig:PXRD}. The \EMCP\ phase dominates this pattern, with minor peaks attributed to \ECP\ and \EMP\ also identified. Lattice parameters of \EMCP\ at room temperature, determined by least-squares fitting of the XRD data, are $a=4.127(2)$ \AA\ and $c=22.29(1)$ \AA, which are slightly higher than those obtained from single-crystal XRD at 150 K.

\clearpage

\subsection{Curie--Weiss analysis for the polycrystalline \EMCP}

As shown in Fig.~\ref{fig:PXRD}, the \EMCP\ phase accounts for over 90\% of the sample composition, with only minor impurity phases (\ECP\ and \EMP) detected, confirming the high phase purity of the powder sample. However, the FM ordering at 30 K originating from the \ECP\ impurity phase, despite its low concentration, dominates the low-temperature magnetic response. This interference precludes the extraction of intrinsic magnetic data for \EMCP\ from the polycrystalline sample. Consequently, we have omitted the $\chi(T)$ curves of polycrystalline \EMCP\ to prevent potential misinterpretation.

\begin{figure}[!htb]
	\includegraphics[width=1.0\textwidth]{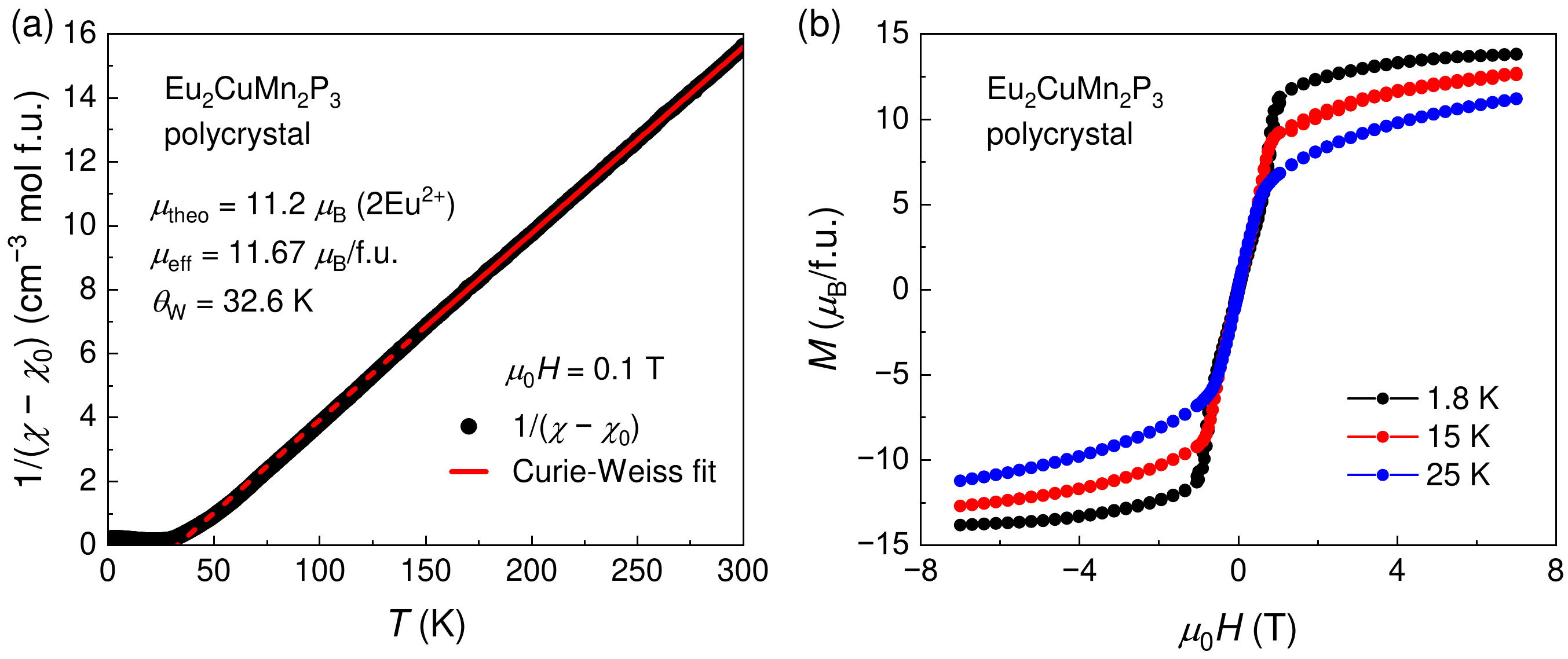}
	\caption{\label{fig:PCmagnetization}
		(a) Inverse susceptibility (black dots) and the Curie--Weiss analysis (red line) of polycrystalline \EMCP\ at 0.1 T. (b)) $M(H)$ curves at 1.8 K (black), 15 K (red), and 25 K (blue).
	}
\end{figure}

Nevertheless, the high-temperature susceptibility data obtained from the polycrystalline sample remain reliable for Curie-Weiss analysis, as presented in Fig.~\ref{fig:PCmagnetization}(a). The Curie--Weiss fit to the data in the 150--300 K range yields an effective moment of 11.67 $\mu_B$/f.u. (corresponding to $C$ = 17.2 K cm$^3$ mol$^{-1}$ f.u.$^{-1}$), a Weiss temperature of 32.6 K, and a temperature-independent contribution $\chi_0=2.21\times10^{-3}$ cm$^3$ mol$^{-1}$ f.u.$^{-1}$. These parameters show excellent agreement with the $\chi_c(T)$ analysis presented in the main text (or see Fig.~\ref{fig:CW_analysis}(b)). Furthermore, the $M(H)$ curves of polycrystalline \EMCP, displayed in Fig.~\ref{fig:PCmagnetization}(b), reveal a saturation moment of 13.8 $\mu_B$/f.u. at 1.8 K, consistent with the values obtained from single-crystal measurements.

\clearpage

\subsection{Supplementary resistivity data for both the single-crystal and polycrystalline \EMCP}

\begin{figure}[!htb]
	\includegraphics[width=1\textwidth]{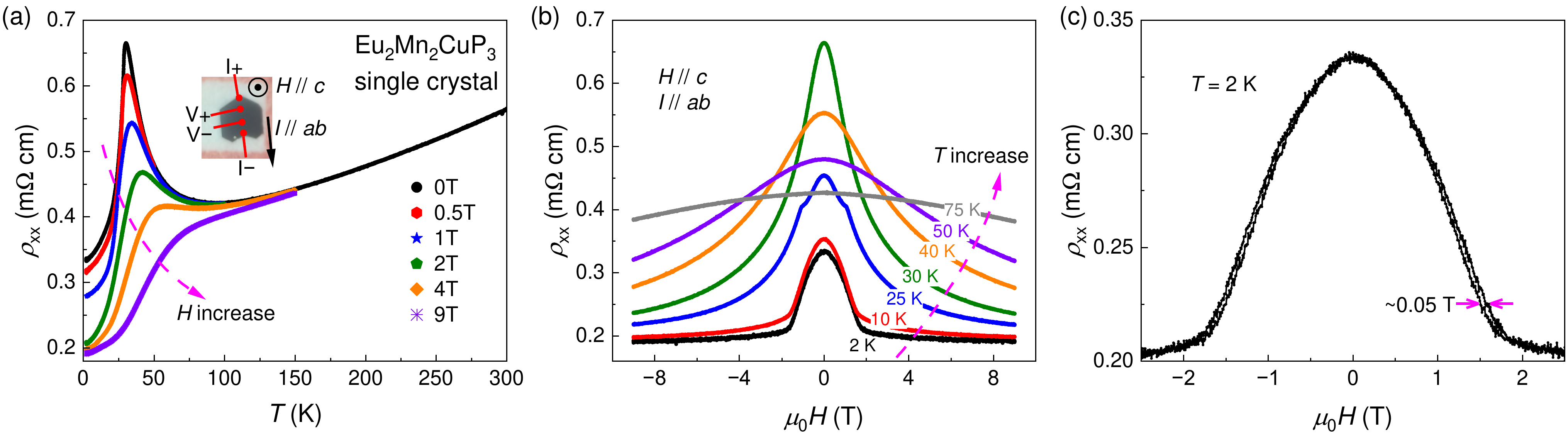}
	\caption{\label{fig:SX_c_Resistivity}
		(a) Temperature-dependent in-plane resistivity of a single-crystal specimen under various out-of-plane magnetic fields ($H\parallel c$). The inset shows a photograph of the \EMCP\ crystal and the electrode configuration. (b) Field-dependent resistivity curves from 2 to 75 K with $H\parallel c$. (c) Slight hysteresis in resistivity at 2 K.
	}
\end{figure}

\begin{figure}[!htb]
	\includegraphics[width=0.8\textwidth]{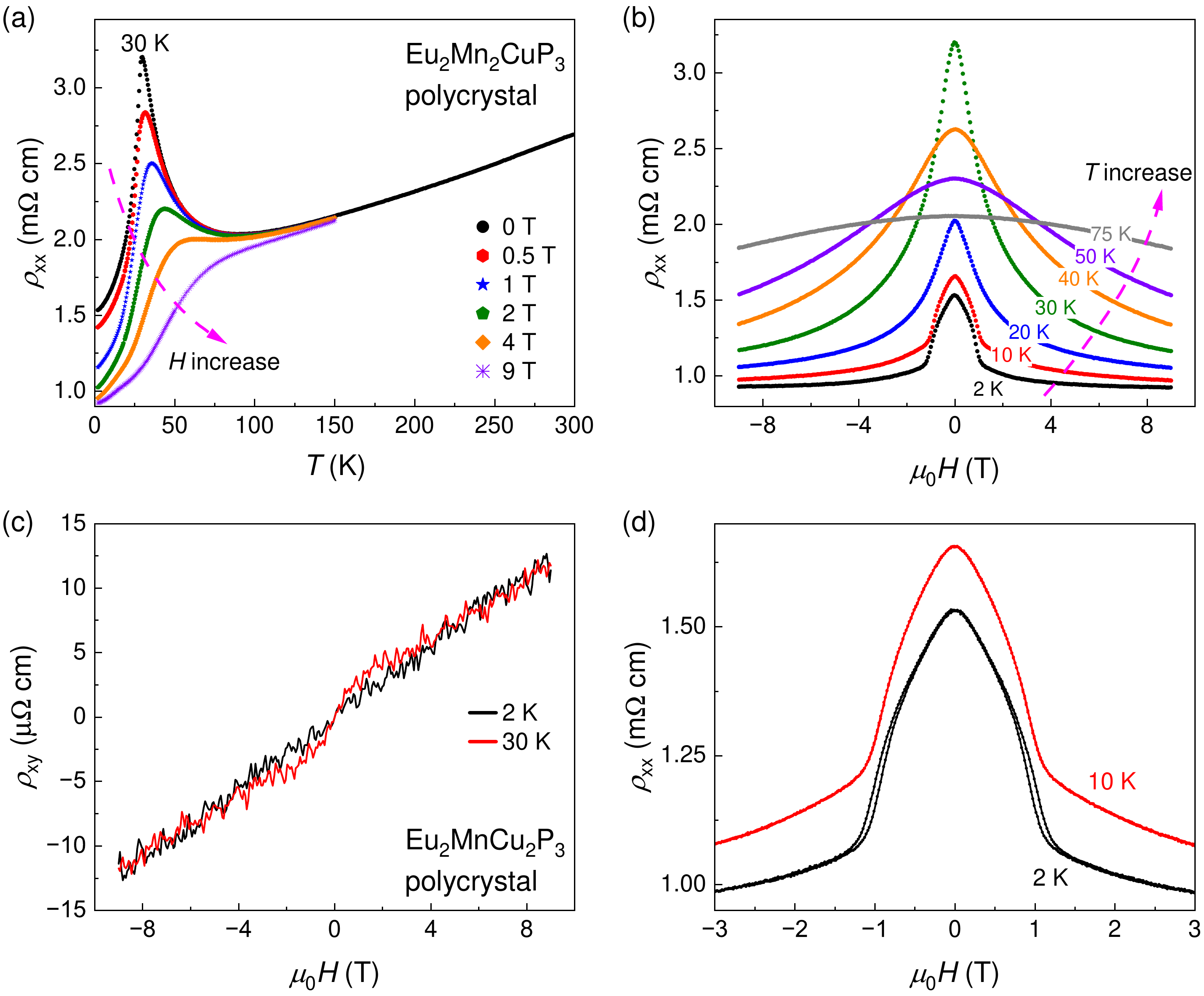}
	\caption{\label{fig:PResistivity}
		(a) Temperature-dependent resistivity of a polycrystalline specimen under various magnetic fields. (b) Field-dependent resistivity curves from 2 to 75 K. (c) Magnetic field dependence of the Hall resistivity at 2 K and 30 K. (d) Slight hysteresis in resistivity at low temperatures.
	}
\end{figure}

The resistivity data for the single crystal ($H\parallel c$) and polycrystalline samples are shown in Fig. \ref{fig:SX_c_Resistivity} and Fig. \ref{fig:PResistivity}, respectively. The in-plane resistivity curves with $H\parallel c$ closely resemble those with $H\perp c$, as shown in the main text, expect for the reduced resistivity hysteresis and an enhanced saturation field of approximately 1.7 T for the resistivity decline. Moreover, the resistivity of the polycrystalline \EMCP\ sample exhibits behavior similar to the anisotropic resistivity observed in single crystals, including the Hall resistivity shown in Fig. \ref{fig:PResistivity}(c). These results suggest that the anisotropy in the resistivity of \EMCP\ is relatively small.

\clearpage

\pagebreak

\subsection{Magnetic energies}

\begin{figure}[!htb]
	\includegraphics[width=0.6\textwidth]{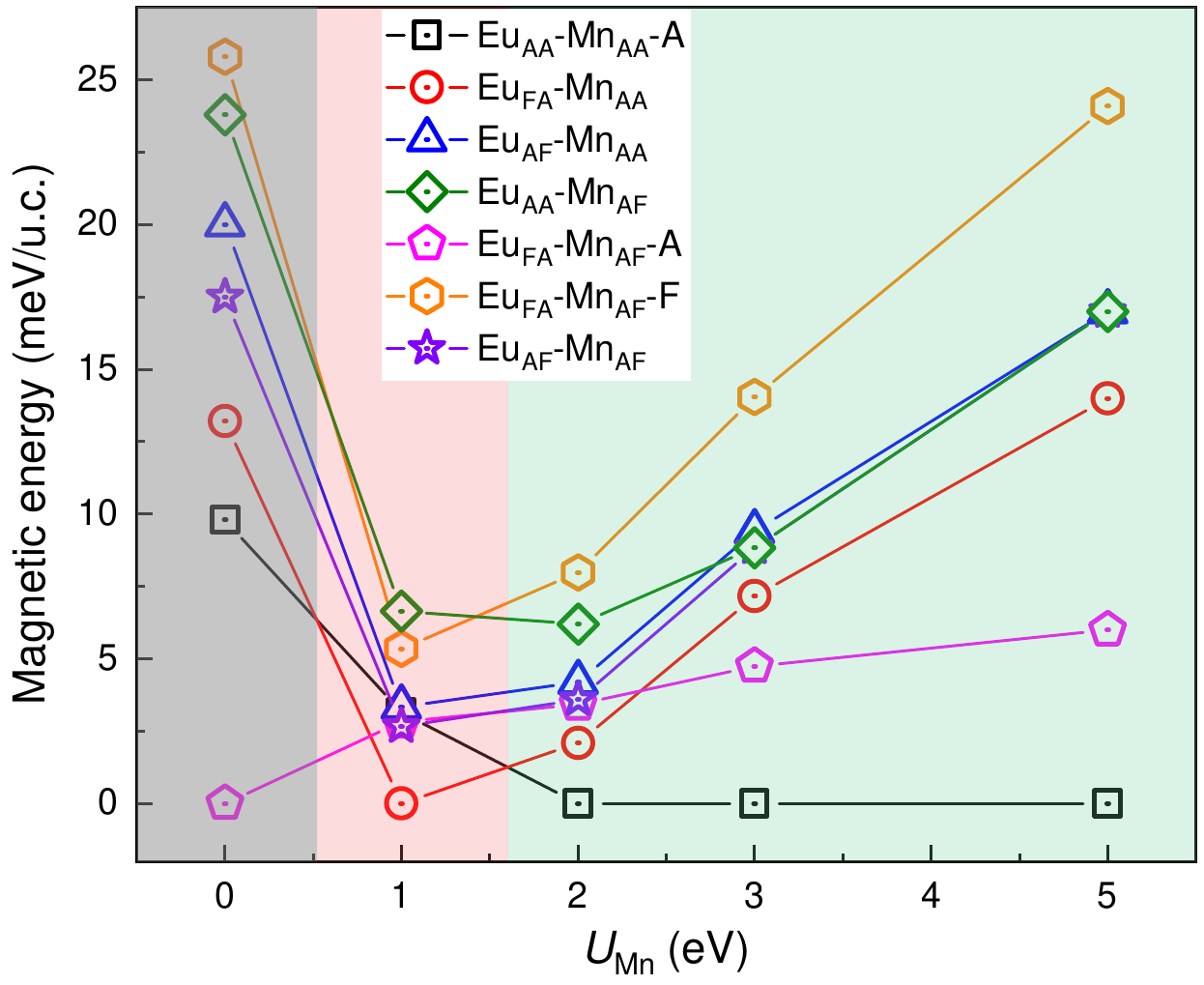}
	\caption{\label{fig:energies}
		The magnetic energies calculated for seven spin configurations with relatively low energy across a range of $U_\mathrm{Mn}$ from 0 eV to 5.0 eV. For each $U_\mathrm{Mn}$, the lowest energy are set to zero.
	}
\end{figure}

Here we test various $U$ values for Mn’s 3$d$ orbitals in \EMCP. Seven spin configurations with relatively low energy were evaluated, and the resulting magnetic energies are plotted in Fig. \ref{fig:energies}. Over a wide range of tested $U_\mathrm{Mn}$ values (about 1.6 eV to 5 eV), the Eu$_\mathrm{AA}$-Mn$_\mathrm{AA}$-A spin structure emerges as the magnetic ground state. We also compared the calculated lattice constants for different  $U_\mathrm{Mn}$ values. Although a large $U_\mathrm{Mn}$ of 5 eV was used in a previous study on \EMP\ (J. Am. Chem. Soc. \textbf{145}, 4527--4533 (2023).), moderate $U_\mathrm{Mn}$ values provide a better match to our experimental lattice constants, as shown in Table \ref{tbl:1}. Additionally, a large $U_\mathrm{Mn}$ of 5 eV introduces a small band gap, as shown in Fig. S6(b), which contradicts the observed metallic behavior for \EMCP. Therefore, we have chosen a moderate $U_\mathrm{Mn}$ of 2 eV for presentation in the main text.

\begin{table}[!htb]
	\caption{Calculated cell parameters of \EMCP\ with different $U_\mathrm{Mn}$ values.}
	
	\label{tbl:1}
	\begin{tabular}{ccccccc}
		\hline
		& Experimental &\multicolumn{5}{c}{$U_\mathrm{Mn}$ (eV)}\\
		\cline{3-7}
		& 150 K & 0  & 1  & 2  & 3  & 5 \\
		\hline
		$a$ (\AA) & 4.121 & 4.039 & 4.108 & 4.146 & 4.172 & 4.204 \\
		$c$ (\AA) & 22.268 & 22.380 & 22.381 & 22.378 & 22.454 & 22.551\\
		$\Delta a$ (\%)\footnote{The percentage difference between the calculated cell parameters and the corresponding experimental value.} & & $-$1.99 & $-$0.32 & 0.61 & 1.24 & 2.0 \\
		$\Delta c$ (\%) &  & 0.50 & 0.51 & 0.49 & 0.84 & 1.3\\
		\hline
	\end{tabular}	
	
\end{table}

\clearpage

%
%
%
~\\

\clearpage

\subsection{Band structures}

\begin{figure}[!htb]
	\includegraphics[width=1.0\textwidth]{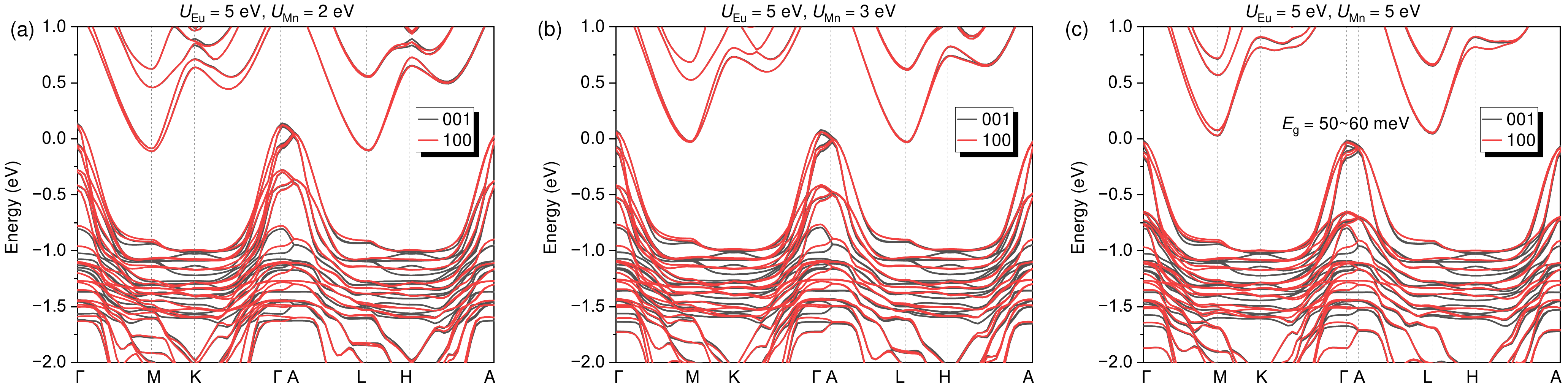}
	\caption{\label{fig:band}
		Band structures calculated with different configurations, considering  spin--orbit coupling (SOC). (a) $U_\mathrm{Eu}$ = 5 eV and $U_\mathrm{Mn}$ = 2 eV. Red and black curves represent the bands calculated with spins aligned along the [001] and [100] directions, respectively. (b) $U_\mathrm{Eu}$ = 5 eV and $U_\mathrm{Mn}$ = 3 eV. (c) $U_\mathrm{Eu}$ = 5 eV and $U_\mathrm{Mn}$ = 5 eV.
	}
\end{figure}

The band structures for \EMCP\ are calculated using different Hubbard $U_\mathrm{Mn}$ values (2.0 eV, 3.0 eV, and 5.0 eV) and spin orientations. As shown in Fig. \ref{fig:band}, the spin orientation has a negligible effect on the band structure. Moderate $U_\mathrm{Mn}$ values of 2.0 eV and 3.0 eV yield semimetallic band structures. However, a larger $U_\mathrm{Mn}$ of 5 eV introduces a small energy gap of 50\textendash60 meV, leading to semiconducting behavior, which is inconsistent with the experimental observations. The band structure calculated with $U_\mathrm{Mn}$ = 2.0 eV is presented in the main text.

